\documentclass[twocolumn,preprintnumbers,amsmath,amssymb,floatfix,superscriptaddress,nofootinbib]{revtex4}
\usepackage{amsmath,hyperref,amssymb}
\usepackage{graphicx}% Include figure files
\usepackage{dcolumn}% Align table columns on decimal point
\usepackage{bm}% bold math
\usepackage{verbatim}
\usepackage{tabularx}
\usepackage{slashed}
\usepackage{cancel}
\usepackage{hyperref}

\numberwithin{equation}{section}

\def\be{\begin{equation}}
\def\ee{\end{equation}}
\def\bea{\begin{eqnarray}}
\def\eea{\end{eqnarray}}
\def\ba#1\ea{\begin{align}#1\end{align}}
\def\bg#1\eg{\begin{gather}#1\end{gather}}
\def\bm#1\em{\begin{multline}#1\end{multline}}
\def\bmd#1\emd{\begin{multlined}#1\end{multlined}}

\renewcommand{\t}{\tilde}
\newcommand{\tr}{\text{tr}}
\newcommand{\sgn}{{\rm{sgn}}}
\newcommand{\SigmaN}{{\Sigma_\text{NFL}}}

\begin{document}

\title{How non-Fermi liquids cure their infrared divergences} 

\author{Jeremias Aguilera Damia}
\affiliation{\small \it Centro At\'omico Bariloche, CNEA and CONICET, Bariloche, R8402AGP, Argentina}
\author{Mario Sol\'is}
\affiliation{\small \it Centro At\'omico Bariloche, CNEA and CONICET, Bariloche, R8402AGP, Argentina}
\author{Gonzalo Torroba}
\affiliation{\small \it Centro At\'omico Bariloche, CNEA and CONICET, Bariloche, R8402AGP, Argentina}

\date{\today}

\begin{abstract}
Non-Fermi liquids in $d=2$ spatial dimensions can arise from coupling a Fermi surface to a gapless boson. At finite temperature, however, the perturbative quantum field theory description breaks down due to infrared divergences. These are caused by virtual static bosonic modes, and afflict both fermionic and bosonic correlators. We show how these divergences are resolved by self-consistent boson and fermion self-energies that resum an infinite class of diagrams and correct the standard Eliashberg equations. Extending a previous approach in $d=3-\epsilon$ dimensions, we find a new ``thermal non-Fermi liquid'' regime that violates the scaling laws of the zero temperature fixed point and dominates over a wide range of scales. We conclude that basic properties of quantum phase transitions and quantum-classical crossovers at finite temperature are modified in crucial ways in systems with soft bosonic fluctuations, and we begin a study of some of the phenomenological consequences.
\end{abstract}

\maketitle

\tableofcontents

%%%%%%%%%%%%%%%%%%%%%%%%%
%%%%%%%%%%%%%%%%%%%%%%%%%
%%%%%%%%%%%%%%%%%%%%%%%%%
%%%%%%%%%%%%%%%%%%%%%%%%%
\section{Introduction}

Finite density quantum field theory (QFT) provides a very promising framework for understanding strongly correlated electronic systems in two spatial dimensions. Already simple models, such as a Fermi surface coupled to a light scalar field, can produce non-Fermi liquid (NFL) behavior. The light scalar field can appear as an emergent soft bosonic mode describing an order parameter for symmetry breaking, or as a gauge field in spin liquids. The Yukawa coupling to the Fermi surface, generically allowed by symmetries, is strongly relevant in $d=2$ dimensions and can give rise to an interacting quantum critical point.

So far, these theories have been most developed at zero temperature, and our goal in this work is to study them at finite temperature. This is important for phenomenological reasons, since most experimental and numerical results of strongly correlated systems are at finite temperature. From a more conceptual viewpoint, we need to understand how quantum and thermal fluctuations compete, and the resulting properties of the ``quantum critical region'' in the phase diagram~\cite{Sachdev}. Furthermore, understanding quantum matter at finite temperature is a central area of research in basic QFT, one that has a long history at zero density (see e.g.~\cite{le2000thermal}).

Finite temperature brings in new infrared divergences that are absent from the zero temperature theory. They originate from the same ingredient that is required to produce a non-Fermi liquid -- exchange of virtual soft bosons. To see how this comes about, let us consider for concreteness a scalar field with a linear dispersion relation (we will analyze other dispersion relations below). At zero temperature, contributions to Feynman diagrams from bosonic internal lines are of the form $\int d^2q d \Omega \frac{1}{q^2+ \Omega^2}$, which is finite at small frequency/momenta. However, at finite temperature the frequency is quantized in terms of Matsubara modes, giving
\be\label{eq:log}
T \sum_n \int \frac{d^2q}{q^2 + (2\pi T n)^2} =T \int \frac{d^2 q}{q^2} + T \sum_{n \neq 0}\int\frac{d^2q}{q^2 + (2\pi T n)^2}\,.
\ee
We see that the exchange of the static $n=0$ mode leads to an infrared logarithmic divergence. We can also understand this by taking a large $T$ limit and dimensionally reducing on the thermal circle; this gives an effective action with a two-dimensional euclidean massless scalar, and a logarithmically divergent Green's function at long distances. New insertions of boson lines in Feynman diagrams will produce more log-divergent powers, leading to a breakdown of perturbation theory. Our aim is to resolve this problem, and understand how it affects quantum criticality.

Thermal divergences in QFT have received a lot of attention at zero density, where they can be cured by nonperturbative effects or by resummation of perturbative corrections.\footnote{The literature is vast; some reviews include~\cite{Gross:1980br, Kapusta:2006pm, le2000thermal}. The title of our work is motivated by~\cite{Jackiw:1980kv}.} In contrast, much less is known about infrared problems at finite density. Work in this area, both in the condensed matter and high energy fields, includes~\cite{PhysRevB.48.7183, Blaizot:1996az, Manuel:2000nh, abanov2003quantum, PhysRevB.80.165116, PhysRevLett.108.186405, PhysRevB.89.155130, PhysRevB.94.195113, Wang:2017kab, Wang:2017teb, klein2020normal, Xu:2020tvb}.

The starting point for our analysis is Ref.~\cite{Wang:2017kab}, so let us briefly summarize its main results. This work focused on NFLs in an $\epsilon$ expansion around $d=3$ dimensions, obtained by coupling a massless overdamped $N \times N$ boson to a Fermi surface of $N$-flavor fermions. The infrared divergences were identified already in the one-loop fermion self-energy, due to exchange of bosons with zero Matsubara frequency (static modes). It was argued that, in order to resolve the thermal divergences, it is necessary to go beyond the usual Eliashberg equations, by resumming an infinite class of perturbative corrections (rainbow diagrams). This resummation is exact when $N \gg 1$. The main consequence of this procedure is that the finite temperature fermion self-energy 
develops a new ``thermal'' NFL contribution $\Sigma_T(\omega_n)$ that comes from exchange of static modes,
\be\label{eq:sigmadecomp}
\Sigma(\omega_n) = \Sigma_T(\omega_n) + \SigmaN(\omega_n)\,.
\ee
The last term here comes from virtual bosons with nonzero Matsubara frequency, has no infrared problems, and recovers the NFL self-energy of the quantum critical point as $T \to 0$. It was found that $\Sigma_T \gg \SigmaN$ over a broad range of frequencies at finite temperature.

In this work we undertake the analysis of the finite temperature dynamics but in $d=2$ spatial dimensions.\footnote{While we were finishing this manuscript, Refs.~\cite{klein2020normal, Xu:2020tvb} appeared, which have some overlaps with the results in Sec.~\ref{subsec:Sigma}.} For reasons of perturbative control, we focus on the large $N$ limit of an overdamped $N \times N$ boson and a Fermi surface of $N$-flavor fermions, coupled through a Yukawa term~\cite{Damia:2019bdx}. However, the lessons on the origin and resolution of IR divergences will apply more generally. The main new effect in decreasing the dimensionality from $d=3$ to $d=2$ is that boson divergences become much more severe, as reviewed in (\ref{eq:log}). As a result, we will have to go beyond~\cite{Wang:2017kab}: we will incorporate quantum corrections to the boson self-energy that have the effect of modifying Landau damping and inducing a self-consistent boson mass. 

We find that the self-consistent boson and fermion self-energies $\Pi$ and $\Sigma$, respectively, are dominated at low frequencies by thermal contributions of the form
\be\label{eq:thermal2}
\Pi(0, q) \sim \lambda_\phi T\;,\;\Sigma_T \sim \sqrt{g^2 T}\,,
\ee
up to logarithmic terms, and where $\lambda_\phi$ and $g$ are the strengths of the 4-boson and Yukawa interactions. The resulting finite temperature QFT is free of IR divergences, depends non-analytically on the original couplings, and has a continuous limit $T \to 0$. These results are presented in Secs.~\ref{sec:origin} and \ref{sec:solution}.

\begin{figure*}[t]
  \centering
  \includegraphics[width=0.9\hsize]{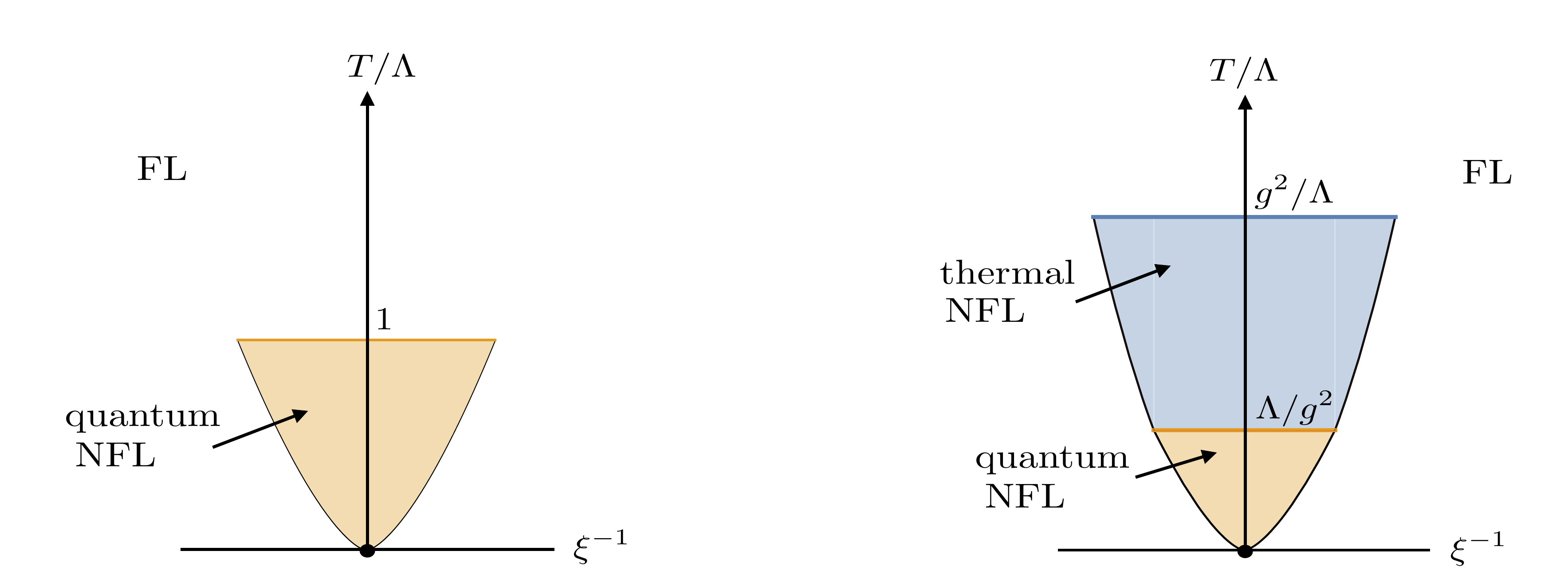}
  \caption{Depiction of phase diagram at finite temperature. Here $\Lambda$ is a dynamical scale that controls the flow to the fixed point and $g$ is the Yukawa coupling; we take $\Lambda \ll g^2$. The left figure ignores thermal effects and represents the traditional quantum critical region; the right figure includes $\Sigma_T$. In this case there appears a thermal NFL regime, which dominates over a wide range of scales, $\Lambda^2/g^2<T<g^2$, and modifies the quantum-classical crossover.}
  \label{fig:qcr1}
\end{figure*}

In the second part of the work (Sec.~\ref{sec:pheno}), we focus on the consequences of these thermal contributions:
\begin{itemize}
\item We show that $\Pi(0,q)$ only affects the static mode and is irrelevant for the higher Matsubara modes. On the other hand, $\Sigma_T$ dominates the fermion dynamics over a wide range of frequencies and temperature.
\item Quantum critical scaling at finite $T$ would imply a typical thermal length $\xi_T \sim T^{-1/z}$, with dynamical exponent $z=3/2$ here. This scaling is violated by $\Sigma_T$, which instead gives $\xi_T \sim (g^2 T)^{-1/2}$.
\item The basic picture of quantum phase transitions and the quantum-classical crossover~\cite{Sachdev} is then dramatically modified by the strong IR effects mediated by the soft boson. We expect the new thermal exponent $\xi_T \sim(g^2 T)^{-1/2}$ to change various observable quantities (thermodynamics and transport).
\end{itemize}
We illustrate these results in Fig.~\ref{fig:qcr1}.

We end the paper with a discussion and future directions in Sec.~\ref{sec:concl}. We also present various detailed calculations in the Appendices.

%%%%%%%%%%%%%%%%%%%%%%%%%
%%%%%%%%%%%%%%%%%%%%%%%%%
%%%%%%%%%%%%%%%%%%%%%%%%%
%%%%%%%%%%%%%%%%%%%%%%%%%
\section{Boson-fermion model at finite temperature}\label{sec:origin}

Although our analysis can be carried out more generally, we will focus on a simple fermion-boson model that features a controlled quantum critical point: a spherical Fermi surface of an $N$-component fermion, coupled to an overdamped $N \times N$ massless boson, with $z_b=3$ dynamical exponent. It is described by the euclidean action $S=S_f+S_b+S_Y$, with\footnote{The integrals are short-hand for $\int_{\omega, p} = \int  \frac{d \omega}{2\pi} \frac{d^2p}{(2\pi)^2}$.}
\bea\label{eq:S0}
S_f &=&- \int_{\omega, p} \psi^\dag_i ( i \omega - \varepsilon_p) \psi^i \nonumber\\
S_b&=& \frac{1}{2} \int_{\Omega, q} \phi_i^j \left(q^2 + M_D^2 \frac{|\Omega|}{q}\right) \phi_j^i \\
S_Y&=&\frac{g}{\sqrt{N}}   \int_{\omega, p}  \int_{\Omega, q} \phi_j^i(\Omega, q) \psi_i^\dag(\omega, p) \psi^j(\omega-\Omega, p-q) \nonumber\,.
\eea
Note that we start with an overdamped scalar field; it can arise as an order parameter, or as the magnetic component of a gauge field. We take $M_D$ as the UV energy scale in the problem, above which a microscopic theory (such as a lattice construction) is needed. We have tuned to zero a possible bare mass term, in order to approach the quantum critical point. We take $N \gg 1$ with fixed $g$.

We restrict to low energies near the Fermi surface, linearizing 
\be
\varepsilon_p = \frac{p^2}{2m} - \mu_F \approx v p_\perp\,,
\ee 
with $\vec p = \hat n (k_F + p_\perp)$. Here $\hat n$ is a unit vector on the Fermi surface, $k_F = \sqrt{2m \mu_F}$, and $v= k_F/m$. From the Yukawa interaction, the boson momentum $\vec q \cdot \hat n$ transverse to the Fermi surface scales like a difference of fermion momenta. Near the position $\hat n$ on the Fermi surface, we shall then decompose $\vec q = q_\perp \hat n+ \vec q_\parallel$. Furthermore, we denote the relative angle by $\cos \theta = \vec q \cdot \hat n /q$.

At large $N$ with fixed $M_D$, the quantum dynamics leads to a critical point that is one-loop exact, recently described in~\cite{Damia:2019bdx}. Corrections to the boson propagator are all suppressed by $1/N$, but there is a leading one-loop fermion self-energy,
\be\label{eq:Sigma1}
\Sigma(\omega) = \frac{g^2}{2\pi \sqrt{3}v}\,\frac{1}{M_D^{2/3}}\,\text{sgn}(\omega)\,|\omega|^{2/3}\,.
\ee
Higher non-planar loop diagrams are also suppressed by $1/N$. Therefore, below the dynamical scale
\be\label{eq:Lambda}
\Lambda = \frac{g^6}{(2\pi  \sqrt{3} \, v)^3 M_D^2 }\,,
\ee
the system flows to a nontrivial NFL with a $z_b=3$ boson ($q^3 \sim M_D^2 |\Omega|$) and $z_f=3/2$ fermion ($p_\perp \sim \Lambda^{1/3} |\omega|^{2/3}$).
\begin{comment}
, and scaling dimensions
\be\label{eq:coord-transf1}
\omega \to e^s \omega\;,\;q_\perp \to e^{\frac{2}{3}s} q_\perp\;,\;q_\parallel \to e^{\frac{s}{3}} q_\parallel\,,
\ee
and
\be\label{eq:field-transf1}
\phi(\omega, q) \to e^{-\frac{4}{3}s} \phi(\omega,q)\,,\, \psi(\omega, q) \to e^{-\frac{7}{6}s} \psi(\omega, q)\,.
\ee
\be\label{eq:coord-transf1}
\omega \to \lambda \omega\;,\;q_\perp \to \lambda^{2/3} q_\perp\;,\;q_\parallel \to \lambda^{1/3} q_\parallel\,,
\ee
and
\be\label{eq:field-transf1}
\phi(\omega, q) \to \lambda^{-4/3} \phi(\omega,q)\,,\, \psi(\omega, q) \to \lambda^{-7/6} \psi(\omega, q)\,.
\ee
\end{comment}
For more details see~\cite{Damia:2019bdx, Torroba:2014gqa, Fitzpatrick:2014cfa, Raghu:2015sna}. We will next consider the fate of this fixed point at finite temperature.

%%%%%%%%%%%%%%%%%%%%%%%%%
%%%%%%%%%%%%%%%%%%%%%%%%%
\subsection{Origin of infrared divergences}

\begin{figure}[h!]
\centering
\includegraphics[width=.8\linewidth]{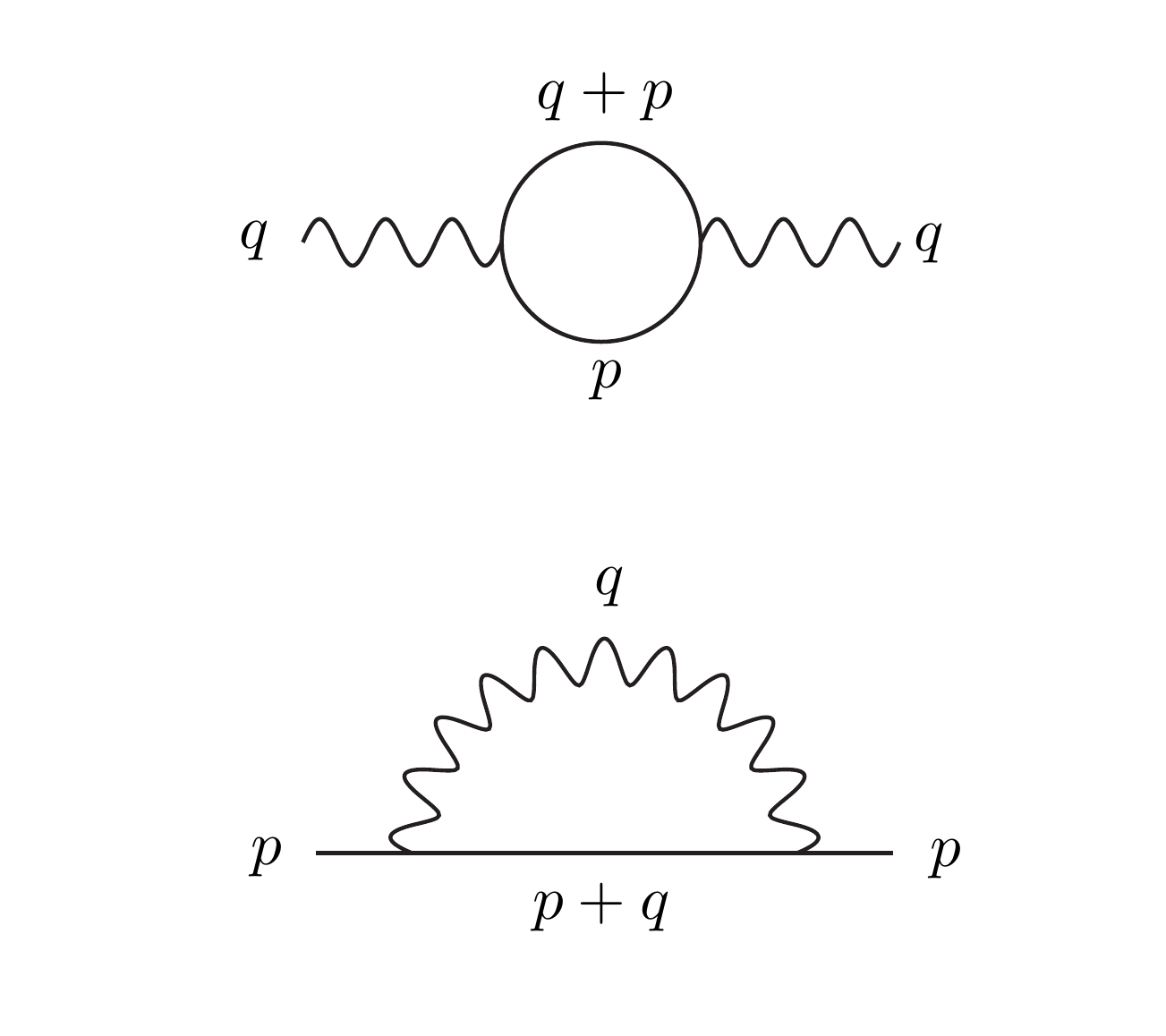}
\caption{\small{One loop boson and fermion self-energies. The wiggly line represents a boson, and the straight line is a fermion; frequencies are omitted to avoid cluttering the diagrams.}}\label{fig:one-loop1}
\end{figure}

At finite temperature, frequency integrals are replaced by Matsubara sums over discrete bosonic ($\Omega_n$) and fermionic ($\omega_n$) frequencies,
\be
\Omega_n = 2\pi T n\;,\; \omega_n= (2n+1) \pi T\,.
\ee
In the action (\ref{eq:S0}), there are two one-loop corrections shown in Fig.~\ref{fig:one-loop1}: a fermion bubble that modifies the bosonic propagator, and a one-loop fermion self-energy. In terms of the tree-level propagators
\be\label{eq:props0}
D(\Omega_n, q) = \frac{1}{q^2 + M_D^2 \frac{|\Omega_n|}{q}}\;,\;G(\omega_n, p) = -\frac{1}{i \omega_n - v p_\perp}\,,
\ee
the two one-loop effects are given, respectively, by
\bea
\Pi(\Omega_n, q)&=& \frac{g^2}{N} T \sum_m\,\int_p G(\omega_m, p) G(\omega_m+\Omega_n, p+q)\quad\; \nonumber\\
i \Sigma(\omega_n) &=& g^2 T \sum_m\,\int_q D(\Omega_m, q)\,G(\omega_n+ \Omega_m,p+q)\,.\;\qquad
\eea

The $z_b=3$ scaling implies that the dependence of the fermion self-energy on the external momentum $p_\perp$ is suppressed by $M_D$, and hence we can neglect it. For this reason, in what follows it will be sufficient to take $p_\perp=0$, so that the momentum lies on the Fermi surface, $\vec p = \hat n k_F$. The independence on $p_\perp$ is required for making the analysis tractable.

Let us consider the Landau damping correction first. A short calculation reproduced in App.~\ref{app:calcs} obtains
\be\label{eq:PIFL}
\Pi(\Omega_n, q) =k_F \frac{g^2}{2\pi v N}\,\frac{|\Omega_n|}{\sqrt{\Omega_n^2+(vq)^2}}\,.
\ee
This is the same as the one-loop Landau damping function at zero temperature. This result is important: it shows that there are no thermal divergences due to fermion bubbles and, crucially for what follows, no boson thermal mass is generated from this diagram, namely $\Pi(0, q)=0$. The original tuning of the bare mass to reach the $T=0$ fixed point is sufficient to keep the boson massless at finite $T$ (at this order). Finally, in the $z=3$ scaling regime, $q^3 \sim M_D^2 |\Omega_n|$, and $\Pi \sim |\Omega_n|/q$. This is of the form of the original kinetic term, but suppressed by $1/N$. In spite of this suppression, we had to go over this derivation to ensure that there is no mass (and no divergences) from this contribution, which otherwise would have been the leading term in the propagator at low frequencies.

Now we come to the fermion self-energy. Denoting the angle between the external fermionic momentum (the position on the Fermi surface) and the internal boson momentum $q$ by $\theta$ we have, more explicitly
\bea
i \Sigma(\omega_n) &=& -g^2 T \sum_m\,\int \frac{d\theta}{2\pi}\frac{q dq}{2\pi}\,D(i \omega_n-i \omega_m, q)\nonumber\\
&&\qquad \times\;\frac{1}{i \omega_m-v q \cos \theta}\,.
\eea
Peforming the angular integral gives
\be\label{eq:Sigma0}
i \Sigma(\omega_n) = i \frac{g^2}{2\pi}T \sum_m\int q dq\,\frac{\text{sgn}(\omega_m)}{\sqrt{\omega_m^2+ (vq)^2}}\,D(i \omega_n-i \omega_m, q)\,.
\ee
We split the Matsubara sum into the $m=n$ mode and the other ones, and use the fact that
\be\label{eq:Dstatic}
D(0, q) = \frac{1}{q^2}\,.
\ee
Then
\bea\label{eq:Sigmaeq2}
\Sigma(\omega_n) &=& \frac{g^2}{2\pi}T\Bigg \lbrace \int  \frac{dq}{q}\,\frac{\text{sgn}(\omega_n)}{\sqrt{\omega_n^2+ (vq)^2}} \\
&+&\sum_{m \neq n}\int q dq\,\frac{\text{sgn}(\omega_m)}{\sqrt{\omega_m^2+ (vq)^2}}\,D(i \omega_n-i \omega_m, q)\Bigg \rbrace\,.\nonumber
\eea
We thus find a logarithmic IR divergence $\int dq/q$ in the $m=n$ contribution, from an exchange of a virtual static boson.

Our derivation so far was for general $D(\Omega, q)$, as long as (\ref{eq:Dstatic}) is satisfied. This emphasizes
that the infrared divergence is quite generic and is associated to a static boson with self-energy $\Pi(0, q) = 0$. This will afflict other NFL models with soft bosons as well. We see that the resolution of this problem requires a mechanism to produce $\Pi(0, q) \neq 0$.

As in~\cite{Wang:2017kab}, the self-energy is a sum a thermal part and a NFL part,
\be\label{eq:SigmaTdef}
\Sigma(\omega_n) = \Sigma_T(\omega_n) + \SigmaN(\omega_n)\,,
\ee
where $\Sigma_T(\omega_n)$ contains the virtual static boson alone, while $\SigmaN(\omega_n)$ includes all the other terms:
\be
\SigmaN(\omega_n) =\frac{g^2}{2\pi}T\sum_{m \neq n}\int q dq\,\frac{\text{sgn}(\omega_m)}{\sqrt{\omega_m^2+ (vq)^2}}\,D(i \omega_n-i \omega_m, q)\,.
\ee
This contribution is expected to be free of IR divergences, and should reproduce the NFL behavior as $T \to 0$. To see this at one loop, we use the tree-level form of the boson propagator (\ref{eq:props0}). The integral over $q$ in the NFL part of (\ref{eq:Sigmaeq2}) is dominated by the bosonic $z=3$ scaling, and gives
\be\label{eq:SigmaNFL1}
\SigmaN(\omega_n) \approx  \frac{g^2 T}{3 \sqrt{3}v} \sum_{m \neq n}\frac{\sgn(\omega_m)}{(M_D^2 |\omega_m - \omega_n|)^{1/3}}\,.
\ee
As discussed in App.~\ref{app:calcs}, this sum can be made in terms of (generalized) Riemann zeta functions, obtaining
\be\label{eq:Rzeta1}
\frac{\SigmaN(\omega_n)}{{\rm sgn}(\omega_n)} \approx  \Lambda^{1/3} (2\pi T)^{2/3}\,\frac{2}{3}\left( \zeta(\frac{1}{3})-\zeta(\frac{1}{3}, |n+\frac12|+\frac12)\right)
\ee
with $\Lambda$ the NFL scale (\ref{eq:Lambda}). Low temperature means $T \ll |\omega_n|$, so that $|n| \gg 1$. In this limit, $\frac{2}{3}\left( \zeta(\frac{1}{3})-\zeta(\frac{1}{3}, |n+\frac12|+\frac12)\right) \approx |n|^{2/3}$, and we indeed recover the NFL result (\ref{eq:Sigma1}). In contrast, so far the additional contribution $\Sigma_T(\omega_n)$ is divergent, and higher loop diagrams will contain additional powers of the IR singularity. So thermal perturbation theory breaks down. 

The last source of divergences at one loop comes from a $\phi^4 $ interaction. This has been neglected so far because it is irrelevant at the fixed point. But it gives a logarithmically divergent contribution to the boson mass, coming from virtual modes with $\Omega_n=0$. This will be discussed in detail in Sec.~\ref{sec:solution}.

At this stage,
it is useful to make make contact with some previous approaches. The form (\ref{eq:SigmaNFL1}) is quite familiar in various contexts, and its generalization
\be\label{eq:gamma-model}
\SigmaN(\omega_n) = g_1^\gamma \pi T\sum_{m \neq n}\frac{\sgn(\omega_m)}{ |\omega_m - \omega_n|^\gamma}
\ee
(with $g_1$ a coupling with dimensions of energy) covers a wide range of NFL behavior. See~\cite{Moon2010, 2019PhRvB..99n4512W, 2019arXiv191201797C} for further details and references. The usual approach is to also take (\ref{eq:gamma-model}) to describe the full self-energy, ignoring $\Sigma_T(\omega_n)$ and discarding the $m=n$ contribution. As explained for instance in~\cite{2019PhRvB..99n4512W}, for calculations of superconductivity this can sometimes be justified by a rescaling procedure (similar to Anderson's theorem). However, this leaves the physical fermionic Green's function undetermined up to an overall scale. As a result, the phenomenological consequences of the NFL theory at finite $T$ cannot be determined, and the question of how infrared divergences are resolved is left unanswered. Here we will seek instead a self-consistent way of determining a finite $\Sigma_T(\omega_n)$.

%%%%%%%%%%%%%%%%%%%%%%%%%
%%%%%%%%%%%%%%%%%%%%%%%%%
\subsection{$\epsilon$-expansion and SD equations}

To obtain a better characterization of the divergence, let us regularize it by deforming the dimensionality to $d=2+\epsilon$, with $\epsilon \ll 1$. Then
\be\label{eq:Sigma-oneloop}
\Sigma_T(\omega_n) \approx \sgn(\omega_n)\, \frac{v^\epsilon T}{2\pi \epsilon}\,\frac{1}{ |\omega_n|^{1-\epsilon}}\,.
\ee
The $1/\epsilon$ factor is equivalent to the log divergence discussed above. On the other hand, the NFL part can be evaluated in terms of zeta functions as in (\ref{eq:Rzeta1})
\bea\label{eq:SigmaNFLep}
\SigmaN(\omega_n) &=&  \frac{g^2}{2\pi \sqrt{3}v}\,\frac{1}{M_D^{\frac{2}{3}(1-\epsilon)}}(2\pi T)^{\frac{2+\epsilon}{3}}\,\sgn(\omega_n)\;\; \\
& \times& \frac{2}{3}\left( \zeta(\frac{1-\epsilon}{3})-\zeta(\frac{1-\epsilon}{3}, |n+\frac{1}{2}|+\frac{1}{2})\right) \qquad\nonumber
\eea
and gives, at low temperatures
\bea
\SigmaN(\omega_n)\approx \Lambda^{\frac{1-\epsilon}{3}} \sgn(\omega_n) |\omega_n|^{\frac{2+\epsilon}{3}}\,.
\eea

\begin{figure}[h!]
\centering
\includegraphics[width=1.1\linewidth]{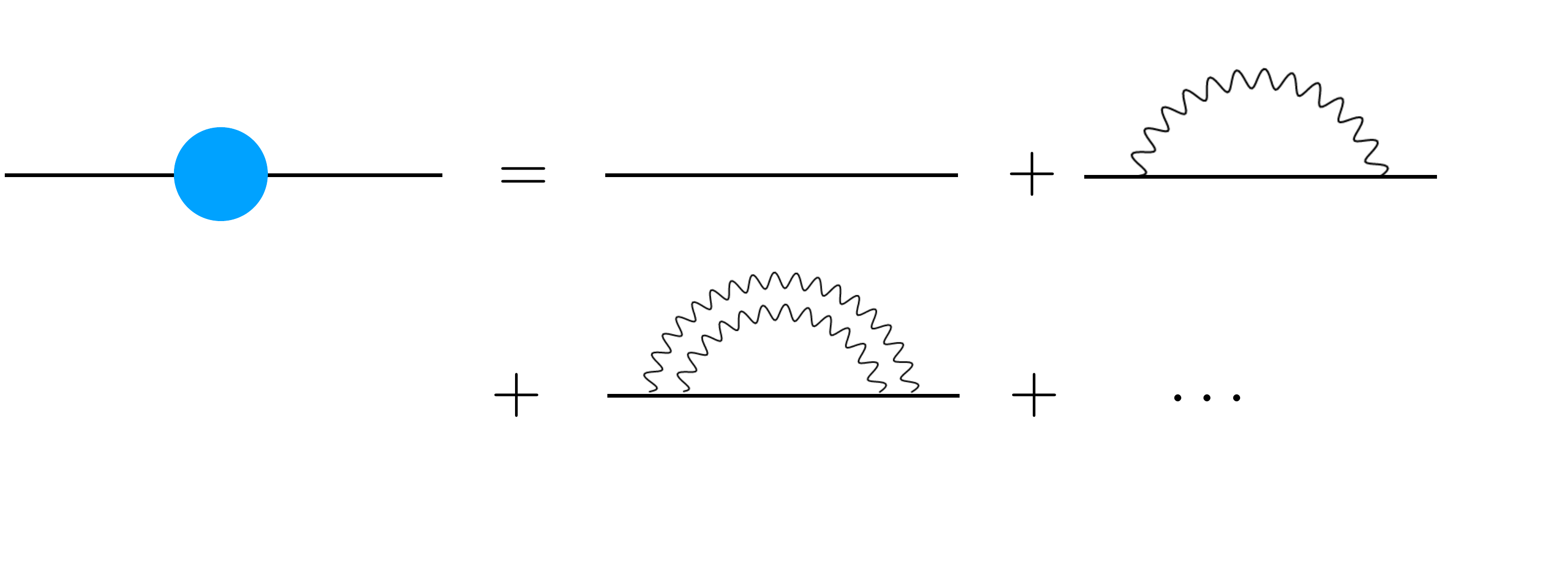}
\caption{\small{Rainbow diagrams for the fermion self-energy. They all have the same leading in $N$ behavior.}}\label{fig:rainbows}
\end{figure}

Ref.~\cite{Wang:2017kab} studied IR divergences in NFLs, in an $\epsilon$ expansion around $d=3$ spatial dimensions. The main result was that resumming an infinite class of perturbative corrections (rainbow diagrams) resolves the thermal singularities. Let us apply the same method here, now in $2+\epsilon$ dimensions.
The rainbow diagrams to be resummed are shown in Fig.~\ref{fig:rainbows}. These diagrams are all leading order in $N$, while other contributions are subleading in $N$.

Introducing the notation
\be\label{eq:Adef}
A(\omega_n) \equiv \omega_n + \Sigma(\omega_n)\,,
\ee
the resummation of rainbow diagrams can be written as a Schwinger-Dyson equation for $\Sigma(\omega_n)$:
\bea\label{eq:SigmaSD1}
\Sigma(\omega_n) &=& \frac{g^2}{2\pi}T\Bigg \lbrace \int  \frac{dq}{q^{1-\epsilon}}\,\frac{\text{sgn}(\omega_n)}{\sqrt{A(\omega_n)^2+ (vq)^2}} \\
&+&\sum_{m \neq n}\int \frac{q^{1+\epsilon} dq}{q^2 + M_D^2 \frac{|\omega_m-\omega_n|}{q}}\,\frac{\text{sgn}(\omega_m)}{\sqrt{A(\omega_m)^2+ (vq)^2}}\Bigg \rbrace\,.\nonumber
\eea
Solving this perturbatively in $g^2$ reproduces the sum of rainbows. The second line in (\ref{eq:SigmaSD1}) is well-approximated by the NFL answer (\ref{eq:SigmaNFLep}).\footnote{This is because in the $z=3$ approximation, this piece is actually independent of $A(\omega_n)$. This is valid as long as $A(\omega) \ll M_D^{2/3}|\omega|^{1/3}$ at low frequencies or temperatures, something that can be checked self-consistently at fixed $\epsilon$.} Performing the momentum integral in the first line, and writing $\Sigma= \Sigma_T + \SigmaN$, we arrive to the following form of the Schwinger-Dyson equation,
\be\label{eq:SigmaT1}
\Sigma_T(\omega_n) = \sgn(\omega_n)\, \frac{v^\epsilon g^2 T}{2\pi \epsilon}\,\frac{1}{ |\omega_n+\Sigma_T(\omega_n) + \SigmaN(\omega_n)|^{1-\epsilon}}\,.
\ee
This is an equation that determines $\Sigma_T(\omega_n)$ explicitly.

Eq.~(\ref{eq:SigmaT1}) is easy to solve numerically. But for our purpose here, it is sufficient to develop intuition about $\Sigma_T(\omega_n)$ by focusing on the low frequency and low temperature behavior, where both $\omega_n$ and $\SigmaN(\omega_n)$ can be neglected in the right hand side of (\ref{eq:SigmaT1}). (In fact, $\SigmaN(\pm \pi T)\approx 0$.) Then we find
\be\label{eq:SigmaTfail}
\Sigma_T(\omega_n) \approx \sgn(\omega_n)\,  \left( \frac{v^\epsilon }{2\pi \epsilon} g^2 T\right)^{\frac{1}{2-\epsilon}}\,.
\ee
This result has several implications.

Taking the limit $\epsilon \to 0$, the resummed $\Sigma_T(\omega_n)$ still diverges. So, in contrast with~\cite{Wang:2017kab}, rainbow resummation has not been enough to cure the IR divergence here. Nevertheless, note that the order of the divergence has decreased from $1/\epsilon$ at one loop, to $1/\epsilon^{1/2}$ at all orders. Another aspect to emphasize in (\ref{eq:SigmaTfail}) is the dependence on $T$: as $\epsilon \to 0$, we find $\Sigma_T \sim (g^2 T)^{1/2}$. In Sec.~\ref{sec:solution}, we will see that this parametric dependence is essentially correct, with the $1/\epsilon^{1/2}$ singularity getting resolved by a self-consistent boson thermal mass.

We can also compare with the results of~\cite{Wang:2017kab} around $d=3$ by taking $\epsilon \to 1$. Replacing $\epsilon=d-2$ and taking $d \to 3$, (\ref{eq:SigmaTfail}) reproduces Eq.~(4.11) in that reference. Furthermore, we see explicitly that there is no divergence in this limit, and rainbow resummation is enough to resolve the thermal divergences.\footnote{But we note that even if (\ref{eq:Sigma-oneloop}) is finite as $\epsilon \to 1$, it still diverges after analytically continuing to real time and taking frequencies $p_0 \to 0$. This problem is solved by rainbow resummation. }

\begin{figure}[h!]
\centering
\includegraphics[width=1.1\linewidth]{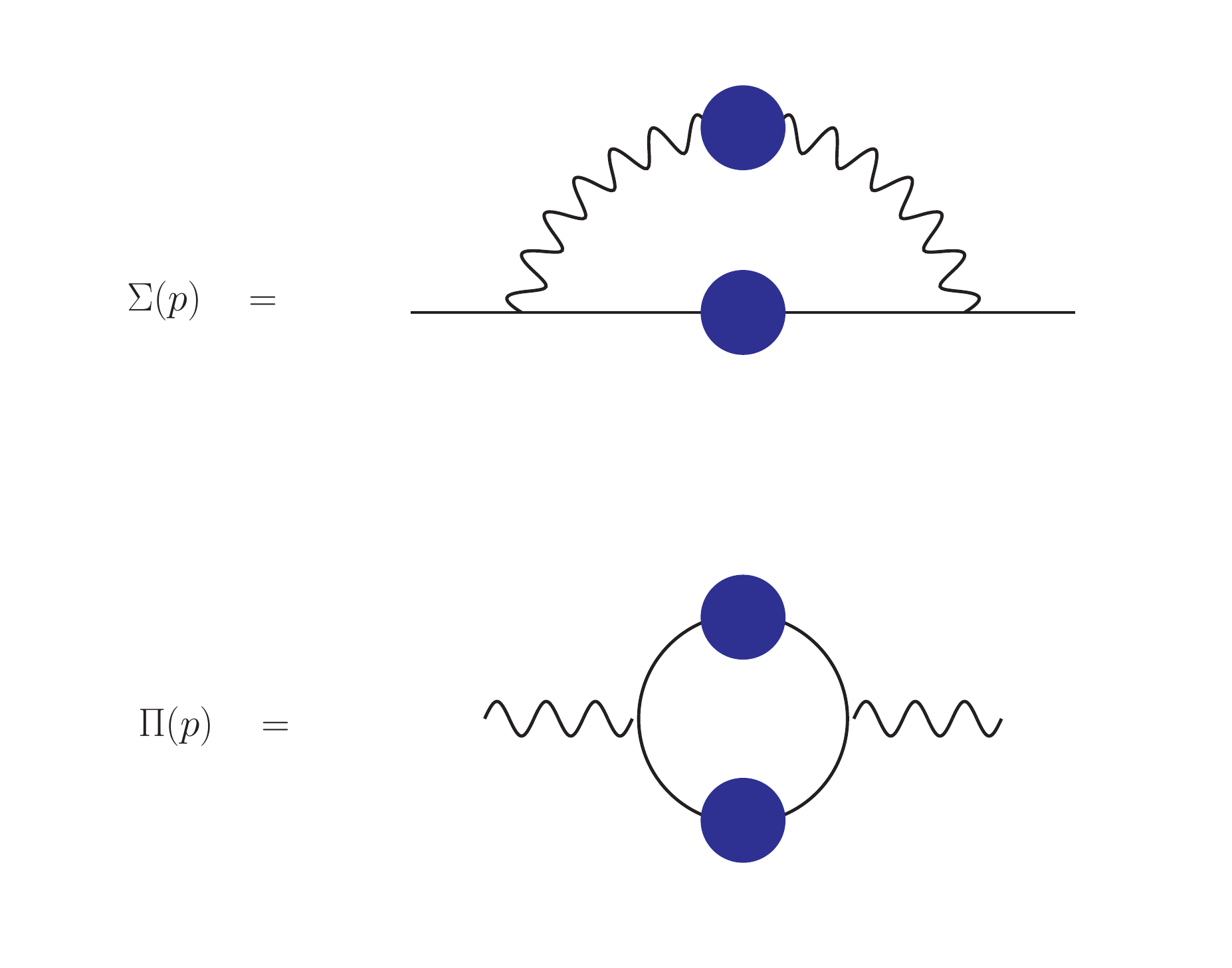}
\caption{\small{Schwinger-Dyson equations for boson and fermion self-energies.}}\label{fig:SD}
\end{figure}

We can try a more general approach by allowing $\Pi(\Omega, q)$ to also adjust self-consistently, solving a coupled system of Schwinger-Dyson equations for the boson and fermion self-energies in the Yukawa model. These are shown in Fig.~\ref{fig:SD}.

Explicitly, the equations for $d=2$ are 
\bea\label{eq:SD}
\Pi(\Omega_n,q) &=&k_F \frac{g^2}{N}T \sum_m \int\frac{dp_\perp}{2\pi}\frac{d\theta}{2\pi}\,\frac{1}{i A(\omega_m)-v p_\perp} \nonumber\\
&\times&\frac{1}{i A(\omega_m+\Omega_n)-v (p_\perp+q \cos \theta)} \nonumber\\
i \Sigma(\omega_n) &=&-g^2 T \sum_m\int  \frac{q dq}{2\pi} \frac{d\theta}{2\pi}\,\frac{1}{i A(\omega_m)- v q \cos \theta}\nonumber\\
&\times&\frac{1}{q^2+M_D^2 \frac{|\omega_m-\omega_n|}{q}+\Pi(\omega_m-\omega_n,q)}\,, \qquad
\eea
with $A(\omega_n)$ defined in (\ref{eq:Adef}). Performing the angular and momentum integrals obtains
\bea\label{eq:SDPi}
\Pi(\Omega_n,q)&=&  k_F \frac{g^2}{v N}\,\sgn(\Omega_n)\,T \\
&\times&\sum_m\,\frac{\Theta(\omega_m+\Omega_n)-\Theta(\omega_m)}{\sqrt{[A(\omega_m+\Omega_n)-A(\omega_m)]^2+ (vq)^2}}\,,\nonumber
\eea
and
\bea\label{eq:SDSigma}
\Sigma(\omega_n) &=&  \frac{g^2}{2\pi}T \sum_m\int q dq\,\frac{\text{sgn}(\omega_m)}{\sqrt{A(\omega_m)^2+ (vq)^2}}\nonumber\\
&\times&\frac{1}{q^2+M_D^2 \frac{|\omega_m-\omega_n|}{q}+\Pi(\omega_m-\omega_n,q)}\,.\qquad
\eea
From (\ref{eq:SDPi}), $\Pi(0, q)=0$ for any $\Sigma(\omega_n)$ -- the two poles in (\ref{eq:SD}) are on the same side. Plugging this into (\ref{eq:SDSigma}) then gives a divergence from the term $m=n$ in the sum. So the boson-fermion system with a Yukawa interaction fails to resolve the infrared problem. 

%%%%%%%%%%%%%%%%%%%%%%%%%
%%%%%%%%%%%%%%%%%%%%%%%%%
%%%%%%%%%%%%%%%%%%%%%%%%%
%%%%%%%%%%%%%%%%%%%%%%%%%
\section{Resolution of the infrared divergences}\label{sec:solution}

So far we have characterized the IR divergences in our finite density QFT, and have argued that they are not rendered finite in the boson-fermion action with just a Yukawa coupling. In this section we will argue that the $\phi^4$ interaction --which we have not discussed so far because it is irrelevant-- actually becomes important for the static boson mode, and provides a natural resolution of thermal divergences through a self-consistent boson mass. We will then add this term to the fermion sector and perform a rainbow resummation in order to obtain a self-consistent fermion self-energy.

%%%%%%%%%%%%%%%%%%%%%%%%%
%%%%%%%%%%%%%%%%%%%%%%%%%
\subsection{A dangerous irrelevant operator in the thermal theory}

Let us focus in more detail on the purely bosonic sector, with the addition of a (single trace) $\phi^4$ interaction,
\bea
S_b&=&\int \frac{d\Omega}{2\pi} \frac{d^2q}{(2\pi)^2} \frac{1}{2}\,\tr\, \left \lbrace\phi_q \left(q^2 + M_D^2 \frac{|\Omega|}{q} \right) \phi_{-q} \right \rbrace \qquad\\
&+& \frac{\lambda_\phi}{8 N} \int \prod_{i=1}^3 \frac{d\Omega_i}{2\pi} \frac{d^2q_i}{(2\pi)^2}\,\tr(\phi_{q_1} \phi_{q_2} \phi_{q_3} \phi_{-q_1-q_2-q_3})\nonumber\,.
\eea
The $\tr( \phi^4)$ interaction is allowed by symmetries (note that it respects the global $SU(N)$ symmetry). It is familiar from the Ising fixed point; if $\phi$ comes from a gauge field, it arises from the kinetic term of the gauge field. It also arises generically in lattice constructions. We keep $\lambda_\phi$ fixed at large $N$, in order to have a finite perturbative expansion. For consistency, we also choose the coupling to be smaller than the UV cutoff, $\lambda_\phi < M_D$.

Here we want to understand the effects of  $\tr( \phi^4)$ on the dynamics of the theory. The coupling has engineering dimensions of an energy scale, $[\lambda_\phi]=1$, the same as $g^2$. But to quantify whether it can lead to important quantum corrections, we need its scaling dimension in the renormalization group sense. Consider first what happens at zero temperature. The quadratic term in $S_b$ is invariant under the scale transformations
\be\label{eq:scalings0}
\Omega \to e^s \Omega\;,\;q \to  e^{\frac{s}{3}} q\;,\;\phi \to e^{-\frac{7}{6}s} \phi\,,
\ee
where the difference in scaling between frequency and momenta reflects the $z_b=3$ dynamical exponent. Under this scaling, the coupling transforms as
\be
\lambda_\phi \to e^{-\frac{s}{3}} \lambda_\phi\,.
\ee
This negative scaling dimension means that $\lambda_\phi$ is irrelevant at the fixed point of the $z_b=3$ damped boson. Quantum corrections at a given energy scale $E$ will come in powers of the dimensionless combination $\lambda_\phi E^{1/3}$, and these vanish in the low energy limit $E \to 0$.

So at zero temperature we can neglect the effects of $\lambda_\phi$ at the fixed point -- and this is indeed what we have done so far. However, the behavior at finite temperature turns out to be quite different. At low energies and momenta, 
\be
 E \ll T, q \ll (2\pi T M_D^2)^{1/3}\,,
 \ee 
 there is a large thermal gap of order $2\pi T$ between the zero mode and the higher Matsubara modes. So the low energy theory contains only the static mode 
\be
\t \phi (q) \equiv T^{1/2} \phi(\Omega_n=0,q)\,,
\ee
and the action becomes
\bea
S_\text{eff}&=&\int \frac{d^2q}{(2\pi)^2} \frac{1}{2}\,\tr\, (\t \phi_q q^2 \t \phi_{-q} ) \qquad\\
&+& \frac{\lambda_\phi T}{8 N} \int \prod_{i=1}^3  \frac{d^2q_i}{(2\pi)^2}\,\tr(\t \phi_{q_1} \t \phi_{q_2}\t \phi_{q_3}\t \phi_{-q_1-q_2-q_3})\nonumber\,.
\eea
This is an euclidean action for a two-dimensional scalar field $\t \phi$, canonically normalized, and with quartic coupling $\lambda_\phi T$.

Under a scale transformation\footnote{We keep the definition of scaling dimension of $q$ from (\ref{eq:scalings0}) so that we can compare with the behavior at $T=0$.} $q \to  e^{\frac{s}{3}} q$ we now have
\be
\t \phi \to e^{-\frac{2}{3}s} \t \phi\;,\;\lambda_\phi T \to  e^{\frac{2}{3}s} \lambda_\phi T\,.
\ee
So in the effective theory for the static mode, the coupling $\lambda_\phi T$ has positive scaling dimension at the gaussian fixed point and becomes relevant. The operator $\phi^4$ then changes from irrelevant in the $2+1$-dimensional theory (high energies/low temperatures) to relevant in the 2-dimensional effective theory (low energies/high temperatures). We call this a ``dangerous irrelevant operator'' in the thermal theory. The name is inspired by a phenomenon sometimes seen in RG flows, whereby an irrelevant operator at high energies becomes relevant at low energies.\footnote{See~\cite{Strassler:2003qg} for a nice review in the relativistic setup.} But we should stress that there are important differences between this and the thermal case where, as we just saw, a dimensional reduction on the thermal circle is operating.

To sum up, the bosonic sector contains an operator $\tr( \phi^4)$, allowed by symmetries, which is irrelevant at the zero temperature fixed point, but becomes relevant at finite temperature. We will next determine its effects on the dynamics of the theory.

%%%%%%%%%%%%%%%%%%%%%%%%%
%%%%%%%%%%%%%%%%%%%%%%%%%
\subsection{Self-consistent boson mass}\label{subsec:mb}

We will carry out our analysis using the low energy effective theory of the static mode, valid up to a cutoff $q< (2\pi M_D^2 T)^{1/3}$. The same result is reproduced including all the Matsubara modes in App.~\ref{app:mb}.
With the addition of the $\phi^4$ interaction, the bosonic sector now has its own IR divergences from exchange of virtual static bosons. 
For instance,  there is a one-loop contribution to the boson mass,
\be
m_b^2 = \lambda_\phi T\,\int \frac{d^2 q}{(2\pi)^2}\,\frac{1}{q^2}\,,
\ee
which diverges logarithmically with an IR cutoff.

\begin{figure}[h!]
\centering
\includegraphics[width=1.1\linewidth]{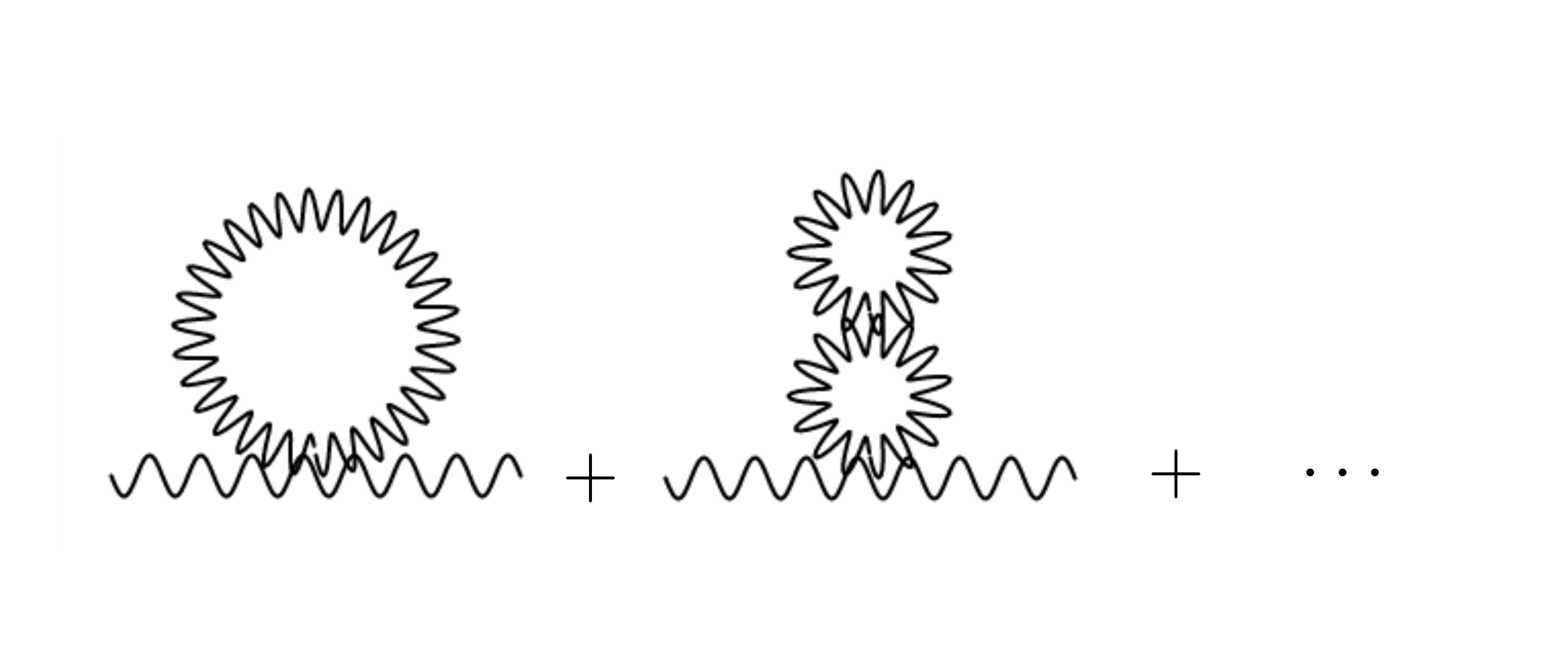}
\caption{\small{Quantum corrections from boson bubble diagrams.}}\label{fig:bubbles}
\end{figure}

The divergence is resolved by summing the boson bubble diagrams, shown in Fig.~\ref{fig:bubbles}. This can be written as a self-consistent equation for a boson mass,
\be\label{eq:mb-self}
m_b^2 = \lambda_\phi T\,\int \frac{d^2 q}{(2\pi)^2}\,\frac{1}{q^2+m_b^2}\,.
\ee
Integrating up to the cutoff of the effective theory gives
\be
m_b^2 = \frac{\lambda_\phi T}{4\pi}\,\log\left( \frac{(2\pi T M_D^2)^{2/3}+m_b^2}{m_b^2}\right)\,.
\ee
At weak coupling,
\be\label{eq:mb}
m_b^2 \approx \frac{\lambda_\phi T}{4\pi}\,\,\log\left(4\pi \frac{(2\pi T M_D^2)^{2/3}}{\lambda_\phi T} \right)\,.
\ee
The argument in the $\log$ is the ratio between the cutoff $(2\pi T M_D^2)^{2/3}$ of the effective theory and the coupling $\lambda_\phi T$.
So by resumming the bubble diagrams we find a finite self-consistent boson mass, which cures the IR divergences. The mass is not analytic in $\lambda_\phi$ at weak coupling; we would then find divergences if we tried to expand it around $\lambda_\phi=0$, which explains the failure of the original perturbative expansion. 

The result (\ref{eq:mb}) used weak coupling, and we need to check whether the flow of the relevant interaction $\lambda_\phi T$ towards strong coupling does not invalidate this. The strong coupling regime is avoided if the mass gap is larger than the scale set by the interaction; this requires
\be\label{eq:no-strong}
m_b^2 \gg \frac{\lambda_\phi T}{4\pi}\,.
\ee
Indeed, the diagrammatic expansion is given in powers of the dimensionless ratio $\lambda_\phi T/(4\pi m_b^2)$. Eq.~(\ref{eq:no-strong}) is guaranteed by the large logarithm in (\ref{eq:mb}), and this is a consequence of $(\lambda_\phi T)/(M_D^2 T)^{2/3} \ll 1$. Therefore, the effective theory of the zero mode never flows to strong coupling, and (\ref{eq:mb}) is valid.\footnote{A related mechanism operates in gauge theories in $2+1$ dimensions; see e.g.~\cite{DHoker:1981bjo}. Another instance of a self-consistent boson mass occurs in the Ising nematic case~\cite{PhysRevB.89.155130}.}

This mass also arises for the higher Matsubara modes, but it is irrelevant on their dynamics. This follows from their $z_b=3$ scaling. Indeed, at a given temperature $T$, they have momenta $q^2 \sim (M_D^2 2\pi T n)^{2/3}$. So the relative size of the mass is
\be\label{eq:highergap}
\frac{m_b^2}{q^2} \lesssim \frac{\lambda_\phi T}{4\pi(M_D^2 2\pi T)^{2/3}} \ll 1\,,
\ee
for all temperatures below the UV cutoff $M_D$, as long as $\lambda_\phi/M_D <1$. The thermal gap then has a negligible effect on all Matsubara modes except for the static one. 

%%%%%%%%%%%%%%%%%%%%%%%%%
%%%%%%%%%%%%%%%%%%%%%%%%%
\subsection{The fermion self-energy}\label{subsec:Sigma}

We will next determine the backreaction of $m_b^2$ on the fermionic sector, and its role in resolving the thermal divergences.

As we argued in the previous section, the thermal mass only has an appreciable effect on the static mode, so it is sufficient to approximate the fermion Schwinger-Dyson equation (\ref{eq:SD}) by
\bea\label{eq:SigmaSD3}
\Sigma(\omega_n) &\approx & \frac{g^2}{2\pi}T\Bigg \lbrace \int  \frac{q dq}{q^2+m_b^2}\,\frac{\text{sgn}(\omega_n)}{\sqrt{A(\omega_n)^2+ (vq)^2}} \\
&+&\sum_{m \neq n}\int \frac{q dq}{q^2 + M_D^2 \frac{|\omega_m-\omega_n|}{q}}\,\frac{\text{sgn}(\omega_m)}{\sqrt{A(\omega_m)^2+ (vq)^2}}\Bigg \rbrace\,.\nonumber
\eea
The first term violates quantum critical scaling and the factorization of transverse and tangential momenta on the Fermi surface~\cite{Wang:2017kab}. It gives a thermal correction to the standard Eliashberg equations that describe the normal state of the fermionic sector.

We will find self-consistently below that $A(\omega_n)$ can be neglected in the second line of (\ref{eq:SigmaSD3}).
Performing the momentum integrals with this approximation, obtains
\be
\Sigma(\omega_n) = \frac{g^2 T}{2\pi}\,\frac{\log \left( \frac{A_n}{v m_b}+\sqrt{\left(\frac{A_n}{v m_b}\right)^2-1}\right)}{\sqrt{A_n^2 - v^2 m_b^2}} + \SigmaN(\omega_n)\,,
\ee
where we recall that $A(\omega_n) = \omega_n + \Sigma(\omega_n)$, $\SigmaN(\omega_n)$ was given in (\ref{eq:SigmaNFLep}). For simplicity we shall focus on $\omega_n>0$ to avoid having to write the sign of the Matsubara frequency in all formulas. This equation determines explicitly the thermal part $\Sigma_T(\omega_n)$ in the decomposition $\Sigma(\omega_n) = \Sigma_T(\omega_n) + \SigmaN(\omega_n)$,
\be\label{eq:SigmaTeq}
\Sigma_T(\omega_n) = \frac{g^2 T}{2\pi}\,\frac{\log \left( \frac{A_n}{v m_b}+\sqrt{\left(\frac{A_n}{v m_b}\right)^2-1}\right)}{\sqrt{A_n^2 - v^2 m_b^2}}\,.
\ee

Let us analyze first the behavior for the first Matsubara mode, $\omega_n = \pm \pi T$, for which $\SigmaN( \pm \pi T )=0$ within the previous approximation. We will work first at sufficiently small temperatures (determined below) so that it is consistent to neglect $\pi T$ compared to $\Sigma_T$ in $A_n$, for the first modes. The equation can be brought to a more convenient form by the change of variables
\be
\Sigma_T (\pi T) = v m_b \, u(\eta)\;,\;\eta \equiv  \frac{g^2 T}{2\pi v^2 m_b^2}\,,
\ee
so that (\ref{eq:SigmaTeq}) becomes
\be\label{eq:u-eq}
u(\eta) = \eta\,\frac{\log \left(u(\eta) + \sqrt{u(\eta)^2-1}\right)}{\sqrt{u(\eta)^2-1}}\,.
\ee
This equation can be solved numerically, and we find the asymptotic behavior
\be
u(\eta) \sim \left \lbrace
\begin{matrix}
\frac{\pi}{2}\eta&,& \text{when} & \eta \ll 1 \\
\sqrt{\frac{1}{2}\eta \log (\eta)}&,& \text{when} & \eta \gg 1 
\end{matrix}
\right. \,.
\ee

For $\eta \gtrsim 1$ (valid except at exponentially low temperatures), our final result for the thermal part of the fermion self-energy is
\be\label{eq:SigmaTsol1}
\Sigma_T (\pi T) \approx \left(\frac{g^2 T}{4\pi} \,\log \left(\frac{g^2 T}{2\pi v^2 m_b^2} \right) \right)^{1/2}\,.
\ee
We conclude that the self-consistent boson mass cuts off the IR divergences, and combining this with rainbow resummation leads to a finite answer $\Sigma_T \sim \sqrt{g^2 T}$. 

The condition to neglect $\omega_n$ compared to $\Sigma_T(\omega_n)$ for $\omega_n = \pi T$ is then $T \ll g^2$. For $T \gg g^2$, we can instead neglect $\Sigma_T$ in the right hand side of (\ref{eq:SigmaTeq}), and it gives $\Sigma_T ( \pi T) \approx g^2/(2\pi^2)$. The contribution of the thermal part to $A(\pi T)$ is then negligible in this regime.

\begin{figure}[h!]
\centering
\includegraphics[width=0.85\linewidth]{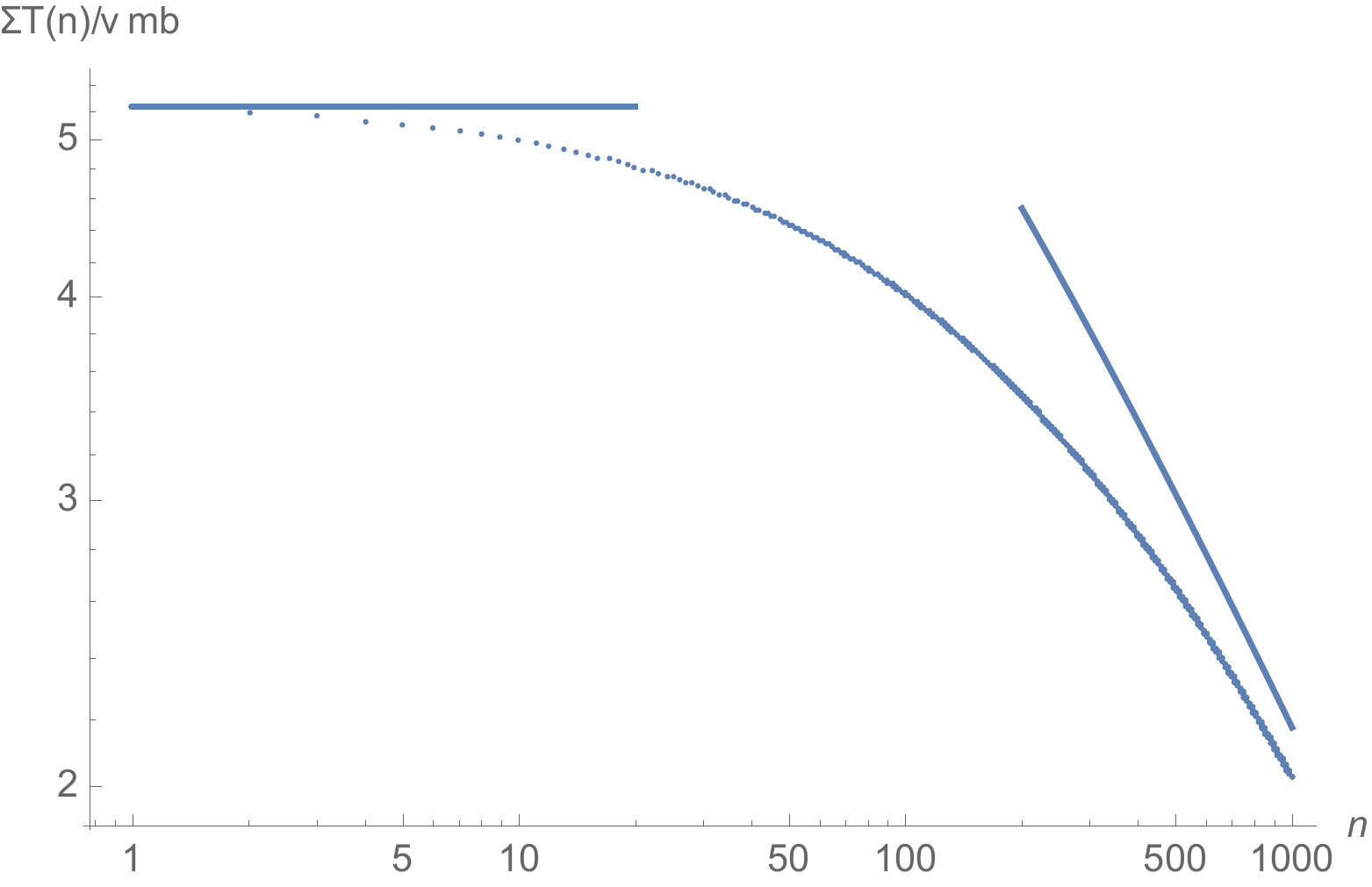}
\caption{\small{Log-log plot of the numerical solution for $\Sigma_T(\omega_n)$, for $T/M_D=10^{-13}, g/M_D=1, \lambda_\phi/M_D=10^{-2}$. The thermal and NFL regimes of (\ref{eq:SigmaT3}) are also shown.}}\label{fig:SigmaT}
\end{figure}

It remains to analyze $\Sigma_T(\omega_n)$ for higher frequencies. For this, we can solve (\ref{eq:SigmaTeq}) numerically, but it is useful to develop intuition about the different regimes. Let us again work at low temperatures (we will fix the temperature window shortly). For the first Matsubara modes, we expect $\Sigma_T(\omega_n)$ to dominate over $\SigmaN(\omega_n)$ in the right hand side of (\ref{eq:SigmaTeq}). Then the solution for $\Sigma_T(\omega_n)$ is approximately the same as in (\ref{eq:SigmaTsol1}). The NFL part starts to compete with the thermal part, $\SigmaN(\omega_n) \sim \Sigma_T(\omega_n)$, at frequencies of order
\be\label{eq:LambdaTdef1}
\Lambda_T(T) \approx \frac{g^{3/2} T^{3/4}}{\Lambda^{1/2}}\,.
\ee
For simplicity of presentation, we avoid writing order one constants and logarithmic corrections, though these are taken into account in the numerical result below. For scales $\Lambda_T < \omega_n < \Lambda$, $\SigmaN(\omega_n)$ will dominate in (\ref{eq:SigmaTeq}); and for $\Lambda< \omega_n < M_D$ (recall that $M_D$ is taken as the highest energy scale in our approach), the Fermi-liquid term $A(\omega_n) \sim \omega_n$ dominates instead. Combining these regimes obtains
\be\label{eq:SigmaT3}
\Sigma_T(\omega_n) = \left \lbrace
\begin{matrix}
\left(\frac{g^2 T}{4\pi} \,\log \left(\frac{g^2 T}{2\pi v^2 m_b^2} \right) \right)^{1/2}\;,\;&&\omega_n < \Lambda_T\\
\frac{g^2T}{2\pi} \frac{\log \left(2 \SigmaN(\omega_n)/v m_b \right)}{\SigmaN(\omega_n)}\;,\; &&\Lambda_T < \omega_n < \Lambda\\
\frac{g^2T}{2\pi} \frac{\log \left(2 \omega_n/v m_b \right)}{\omega_n}\;,\; &&\Lambda < \omega_n < M_D
\end{matrix}
\right.\,.
\ee
The existence of these three regimes requires $\Lambda_T(T) < \Lambda$ or, equivalently, $T/\Lambda < \Lambda/g^2$. We show the numerical solution and the first two analytic behaviors in Fig.~\ref{fig:SigmaT}.

 For larger temperatures, we find that the intermediate NFL regime is absent -- this is an important dynamical consequence of the thermal corrections, which we explore more in Sec.~\ref{sec:pheno}. In this case,
 \be\label{eq:SigmaT2}
\Sigma_T(\omega_n) = \left \lbrace
\begin{matrix}
\left(\frac{g^2 T}{4\pi} \,\log \left(\frac{g^2 T}{2\pi v^2 m_b^2} \right) \right)^{1/2}\;,\;&&\omega_n < \Lambda_T'\\
\frac{g^2T}{2\pi} \frac{\log \left(2 \omega_n/v m_b \right)}{\omega_n}\;,\; &&\Lambda_T' < \omega_n < M_D
\end{matrix}
\right.\,,
\ee
with transition scale
\be\label{eq:LambdaTp}
\Lambda_T'(T) \approx \sqrt{g^2 T}\,.
\ee
The second regime in (\ref{eq:SigmaT2}) was also recently found in~\cite{Xu:2020tvb}, where it was argued to be relevant for matching quantum Monte Carlo results~\cite{Xu:2017dtl}.

We should also verify our assumption that $A(\omega_m) \ll v q$ in the second line of (\ref{eq:SigmaSD3}), which allowed us to replace that term by $\SigmaN(\omega_n)$ of (\ref{eq:Rzeta1}). The only new possibility here is that the thermal term $\Sigma_T$ violates the $z=3$ scaling form of the $T=0$ result. So it is sufficient to approximate $A(\omega_n) \approx \Sigma_T$. Changing variables to $v q/\Sigma_T$, it is not hard to see that the $z=3$ result will continue to hold as long as
\be
\frac{M_D^2 |\omega_m-\omega_n|}{(\Sigma_T/v)^3} \sim \frac{M_D^{3/2}}{g^3}\,\frac{M_D^{1/2}}{T^{1/2}} \gg1\,.
\ee
This is always satisfied in our theory. See Sec.~\ref{sec:pheno} for more details on the relations obeyed by these energy scales.

To end,
it is important to stress that $\Sigma_T(\omega_n)$ is not determined by the $T=0$ dynamics, and violates the scaling laws of the quantum critical point. These would require $\Sigma_T \sim T^{2/3}$, but instead we find $\Sigma_T \sim T^{1/2}$.

%%%%%%%%%%%%%%%%%%%%%%%%%
%%%%%%%%%%%%%%%%%%%%%%%%%
%%%%%%%%%%%%%%%%%%%%%%%%%
%%%%%%%%%%%%%%%%%%%%%%%%%
\section{Phenomenological consequences}\label{sec:pheno}

With a view towards phenomenological applications, in this section we will put together our previous results, and study their consequences for quantum criticality at finite temperature and frequencies. We will also comment on BCS pairing interactions and superconductivity.

At zero temperature, the dimensionful scales are $M_D$ (the UV cutoff), the relevant coupling $g$ that is responsible for the quantum critical point, and the strength $\lambda_\phi$ of the $\phi^4$ interaction. We will choose to work at weak coupling $g^2 \ll M_D$. Then the NFL scale below which the system flows to the quantum critical point is (see (\ref{eq:Lambda}))
\be\label{eq:Lambda2}
\Lambda \approx \left(\frac{g^2}{M_D} \right)^2 g^2
\ee
and we have
\be\label{eq:scales}
\Lambda \ll g^2 \ll M_D\,.
\ee
Furthermore, since $\lambda_\phi$ is irrelevant at the fixed point, it is natural to take it to be small, $\lambda_\phi \ll g^2$. We will now discuss the rich dynamics that ensues from turning on finite temperature, and the competition between quantum and thermal effects.

%%%%%%%%%%%%%%%%%%%%%%%%%
%%%%%%%%%%%%%%%%%%%%%%%%%
\subsection{Quantum and thermal dynamics}\label{subsec:QT}

We discuss the behavior of the fermionic $A(\omega_n) = \omega_n + \Sigma_T(\omega_n)+ \SigmaN(\omega_n)$ which, according to Sec.~\ref{subsec:Sigma}, depends strongly on $T$ and $\omega_n$. We will focus on the parametric dependence, avoiding order one numerical factors for clarity of presentation. We distinguish four ranges of temperatures:

\textit{1)} $T=0$:
\be
A(\omega) \approx \left \lbrace
\begin{matrix}
\Lambda^{1/3} \omega^{2/3}\;,\;&& \omega< \Lambda \\
\omega\;,\;&& \Lambda< \omega < M_D
\end{matrix}
\right.
\ee

\textit{2)} $0<T< (\Lambda/g^2)\Lambda$:
\be
A(\omega_n) \approx \left \lbrace
\begin{matrix}
\sqrt{g^2 T}\;,\;&& \omega_n < \Lambda_T \\
\Lambda^{1/3} \omega_n^{2/3}\;,\;&& \Lambda_T<\omega_n< \Lambda \\
\omega_n\;,\;&& \Lambda< \omega_n < M_D
\end{matrix}
\right.
\ee
We recall that
\be
\frac{\Lambda_T(T)}{\Lambda } \approx \left( \frac{g^2}{\Lambda}\right)^{3/4} \left( \frac{T}{\Lambda}\right)^{3/4}\,.
\ee
So in the current range of temperatures, $\Lambda_T(T)< \Lambda$, and the three different behaviors arise -- thermal NFL, quantum NFL, and Fermi liquid. A large number of Matsubara modes belong to each of these regimes. As the temperature increases, the scale separation between $\Lambda_T$ and $\Lambda$ decreases, eventually shrinking to zero for $T/\Lambda \approx \Lambda/g^2$.

\textit{3)} $(\Lambda/g^2) \Lambda< T< g^2$:
\be
A(\omega_n) \approx \left \lbrace
\begin{matrix}
\sqrt{g^2 T}\;,\;&& \omega_n< \Lambda_T' \\
\omega_n\;,\;&& \Lambda_T'< \omega < M_D
\end{matrix}
\right.\,.
\ee
In this range the quantum NFL disappears, and the fermionic behavior transitions from thermal NFL directly into the Fermi liquid regime, at a scale
\be
\frac{\Lambda_T'(T)}{\Lambda} \approx \left(\frac{g^2}{\Lambda} \right)^{1/2}  \left( \frac{T}{\Lambda}\right)^{1/2}\,.
\ee
Note that $\Lambda_T = \Lambda_T'$ at the boundary $T/\Lambda= \Lambda/g^2$ between 2) and 3). The disappearance of the quantum NFL regime is a consequence of the strong IR divergences. Indeed, ignoring $\Sigma_T$, the NFL behavior disappears at $T \approx \Lambda$. But here this happens instead for $T \approx (\Lambda/g^2) \Lambda \ll \Lambda$.

\textit{4)} $g^2< T< M_D$:
\be
A(\omega_n) \approx \omega_n\;,\;\omega_n<M_D\,.
\ee
The totality of the fermionic modes present a linear dispersion, thus being a standard Fermi liquid (FL) regime.

\begin{figure}[h!]
\centering
\includegraphics[width=1.3\linewidth]{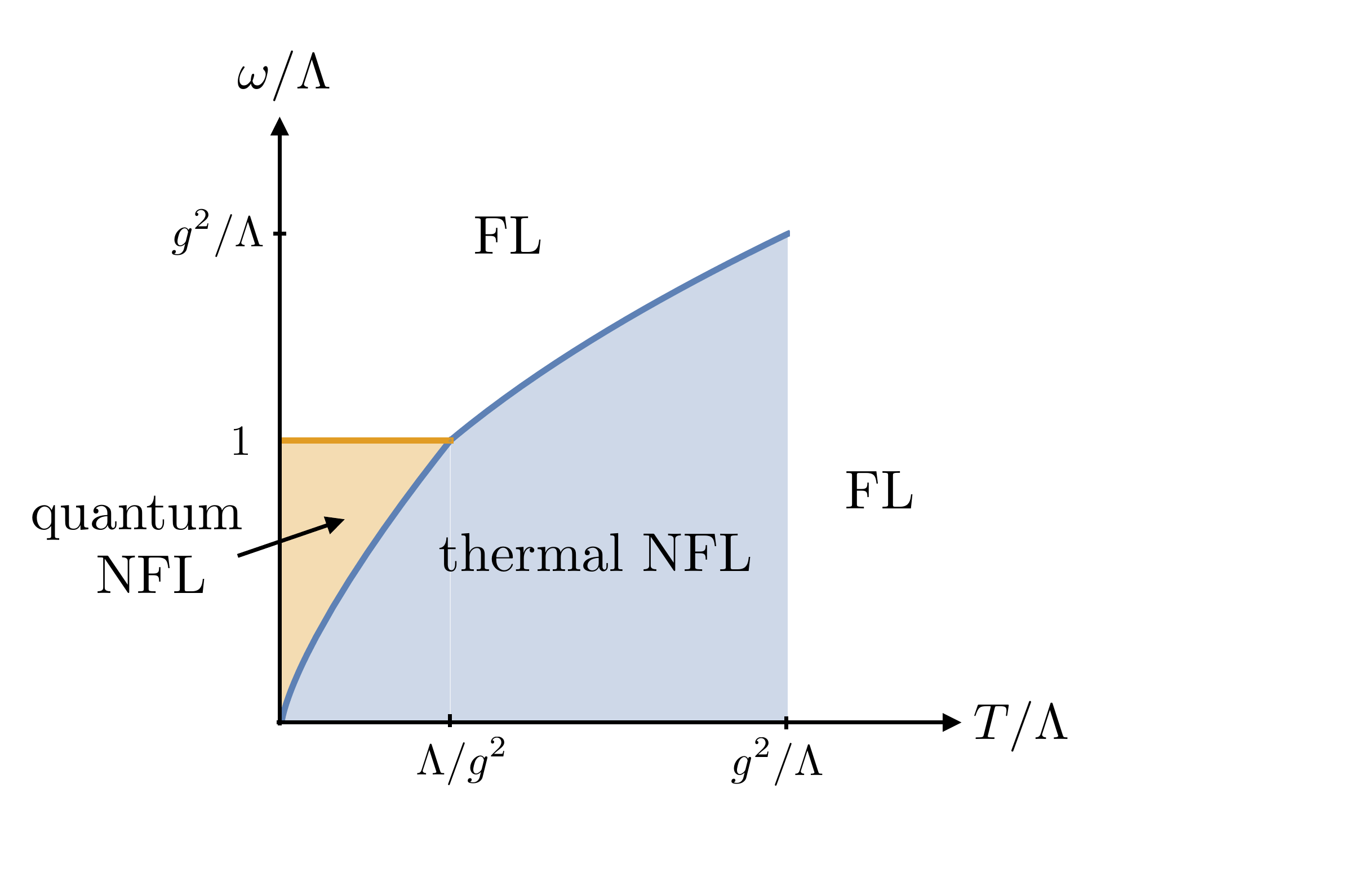}
\caption{\small{Behavior of the fermionic $A(\omega_n)$ in $(T, \omega_n)$. The two curves delimiting the thermal NFL region from above are $\Lambda_T(T)/\Lambda$ and $\Lambda_T'(T)/\Lambda$.}}\label{fig:diagram}
\end{figure}

We combine these results in Fig.~\ref{fig:diagram}. This diagram summarizes the interplay between quantum, thermal and Fermi-liquid effects. For instance, fixing a frequency $\pi T<\omega< \Lambda$ and increasing the temperature, gives quantum NFL, thermal NFL, and FL behavior. The quantum critical region, defined as the regime controlled by the $T=0$ fixed point scaling, then extends up to a temperature $\Lambda_T(T) \sim \omega$. At higher temperatures, a different type of NFL behavior emerges -- the thermal regime associated to nonperturbative effects from static modes. This is not controlled by the quantum critical point scaling. Eventually, for $T \sim g^2$, a standard FL regime appears. A plot of this type was given in Fig.~\ref{fig:qcr1}.
Similar conclusions were reached in~\cite{Wang:2017kab} for $d=3-\epsilon$.

It only remains to check whether the Green's function of the boson is modified in these regimes. Taking into account the restrictions from the Heaviside functions on the Matsubara sum in (\ref{eq:SDPi}), obtains
\bea
\Pi(\Omega_m, q) &=& \frac{k_F g^2}{v N}T\\
&\times&\sum_{n=1}^m\,\frac{1}{\sqrt{(|A(\omega_n+\Omega_m)|+|A(\omega_n)|)^2+(vq)^2}}\nonumber\,.
\eea
Neglecting $A(\omega)$ compared to $vq$ leads to a Landau damping for a $z_b=3$ boson. But now we want to determine whether effects from $A(\omega)$ here can be important.

For this, let us focus on the first regime above, with $T/\Lambda< \Lambda/g^2$, so that $\Lambda_T < \Lambda$. The Matsubara sum is constrained by the external frequency $\Omega_m$. In order to check whether the new thermal part $\Sigma_T$ can modify the boson self-energy, we can consider $\Omega_m < \Lambda_T$; then for all the Matsubara modes in the sum $A(\omega) \approx \Sigma_T$ and we have
\be
\Pi(\Omega_m, q) =\frac{k_F g^2}{2\pi v N}\,\frac{|\Omega_m|}{\sqrt{(2 \Sigma_T)^2 + (vq)^2}}\,.
\ee
Taking $q \to 0$ gives 
\be\label{eq:Pi0}
\Pi(\Omega_m, 0) =\frac{k_F g^2}{4\pi v N \Sigma_T} |\Omega_m|\,.
\ee 
This is an interesting modification from the Fermi-liquid result (\ref{eq:PIFL}), which instead gives a mass term in this limit.
A self-energy that is linear in frequency could give a $z_b=2$ boson; however, it is not hard to check that this does not happen near the mass-shell, $q^2 \sim \Pi(\Omega, q)$, since there $(vq) \gg \Sigma_T$. Hence on-shell bosons do not obey a $z_b=2$ scaling, something also observed in~\cite{PhysRevB.94.195113}.
Nevertheless, there can be off-shell processes for which (\ref{eq:Pi0}) matters, as discussed recently in~\cite{klein2020normal} in connection with the Monte Carlo results~\cite{Schattner2016, Xu:2017dtl, 2017PNAS..114.4905L, Berg2019, liu2019itinerant}.

We conclude that, while thermal contributions modify the fermion dynamics in important ways, their effects on the boson are negligible on-shell, besides the thermal boson mass for the static mode. Recalling (\ref{eq:highergap}), for most of the Matsubara modes we obtain a nearly gapless $z_b=3$ Green's function at finite temperature and at the nonperturbative order that we have worked.

%%%%%%%%%%%%%%%%%%%%%%%%%
%%%%%%%%%%%%%%%%%%%%%%%%%
\subsection{BCS interactions and superconductivity}\label{subsec:SC}

One of the main consequences of our results is that, due to the self-consistent boson mass and the resolution of IR divergences, the dynamics is continuous as $T \to 0$. This is generally expected in QFT, and we have shown how it works out in detail at finite density. On the other hand, so far we have not included the BCS 4-Fermi interaction, which becomes relevant at one loop and could induce a superconducting instability. In fact, the formation of a superconducting gap can also resolve IR divergences and hence can compete with the mechanism we have presented in previous sections. We will see, however, that this does not occur as long as $N>8$.

For this purpose, we shall study the interplay between pairing interactions and typical effects of incoherence due to NFL dynamics, establishing the absence of pairing instabilities at large $N$. We do this at $T=0$, postponing the treatment at finite temperature to a future work. Because of continuity between the zero and finite temperature theories, we expect that our results will also hold at finite temperature.\footnote{This was already the case in $d=3-\epsilon$~\cite{Wang:2017teb}. However, one can also modify the Schwinger-Dyson equations as in~\cite{2019PhRvB..99n4512W,2019arXiv191201797C}, leading to a discontinuity between $T=0$ and finite $T$ dynamics due to an enhanced role of first Matsubara frequencies. This, however, does not occur in the QFT models we consider.}

We extend the approach developed in~\cite{Raghu:2015sna,Wang:2016hir} for $d=3-\epsilon$ to $d=2$. Our conclusions will be qualitatively similar: we will find a critical value $N_c$, such that for $N> N_c$ the system does not superconduct but instead develops critical BCS interactions. For $N< N_c$ a superconducting instability develops; both regimes are separated by an infinite order BKT-type transition as $N \to N_c$.

Pairing interactions are driven by the standard marginal 4-fermion BCS interaction. Since the leading contribution of the boson exchange to this interaction arises in the $s$-channel, we will focus on this case,
\bea
S_{BCS}=-\frac{\lambda}{4k_FN} \int &&(\prod_{l=1}^3 d\omega_l) dq_\perp dp_\perp dp_\perp^\prime d\vec{n}  \times \nonumber\\ 
&&\times 
\psi_i^{\dagger}(3) \psi^j(1) \psi_j^\dagger(4) \psi^i(2) \qquad
\eea
where $p=(k_F+p_\perp)\vec{n}$, $p^\prime=(k_F+p^\prime_\perp)\vec{n}$, $q=q_\perp\vec{n}$ and
\bea
\psi(1)= \psi(\omega_1, p) \quad &,& \quad \psi(2)=\psi(\omega_2,-p) \nonumber\\
\psi(3)= \psi(\omega_3, p+q) \quad &,& \quad \psi(4)=\psi(\omega_4,-p-q) \nonumber
\eea
with $\omega_4=\omega_1+\omega_2-\omega_3$.

There are two ways of approaching the problem. The first uses the renormalization group beta function for the BCS interaction. As analyzed in~\cite{Raghu:2015sna}, it has three terms: a constant tree-level contribution due to boson exchange~\cite{Son:1998uk}, a linear in $\lambda$ term due to the fermion anomalous dimension~\cite{Raghu:2015sna}, and the one-loop $\lambda^2$ contribution originally obtained in~\cite{Shankar, Polchinski},
\be
\frac{d\lambda}{d\log\mu}= -\frac{8\pi}{3}\alpha+\alpha \lambda -\frac{\lambda^2}{4\pi N} \label{BCSRG}
\ee
with $\alpha=1/3$ the fixed point value of the coupling $\alpha \sim g^2/v$~\cite{Damia:2019bdx}. This beta function is a concrete instance of the competition between incoherence and pairing fluctuations. The RG flow $\lambda(\mu)$ is determined by the discriminant of the right hand side of (\ref{BCSRG}). For $N>8$, the right hand side vanishes at two real values
\be
\lambda_\pm = \frac{2\pi}{3} N\left( 1 \pm \sqrt{1-8/N}\right)\,.
\ee
These two roots collide as $N \to 8$, and disappear into the complex plane for $N<8$.

Therefore, as long as $N$ is larger than the critical value
\be
N_c=8 \label{criticalN}\,,
\ee
the BCS coupling flows to a stable IR fixed point
\be\label{fixedBCS}
\lambda_* = \frac{2\pi}{3} N\left( 1 - \sqrt{1-8/N}\right)\,.
\ee
This intriguing quantum state has critical pairing interactions, and was analyzed in~\cite{Raghu:2015sna} when $d=3-\epsilon$. Similar considerations apply in our current $d=2$ setup. We see then that at large $N$ and up to $N=8$, quantum criticality wins over superconductivity, preempting the pairing instability and giving rise to a critical point including BCS interactions. The quantum and thermal dynamics obtained in previous sections is then expected to apply for $N>N_c$.

The IR and UV fixed points $\lambda_\pm$ annihilate for $N=N_c$, and superconductivity develops. On general grounds~\cite{Kaplan:2009kr}, this is expected to be an infinite order Berezinskii-Kosterlitz-Thouless type (BKT) transition. This can be seen by computing the correlation length $\xi$ in the superconducting state as $N \to N_c$:
\be\label{eq:xi}
\frac{\xi^{-1} }{M_D }\approx\,\exp \left(\int_{\lambda_{UV}}^{\lambda_{IR}} \frac{d\lambda}{d\lambda/d\log \mu} \right) \approx \,\exp\left(-\frac{6\pi}{\sqrt{8/N-1}} \right)\,.
\ee
This vanishes as $N \to 8$, but is infinitely differentiable there -- a BKT transition.
A detailed analysis of the solution to \eqref{BCSRG} is presented in Appendix~\ref{runningBCS}. 

Another possibility is to use the Schwinger-Dyson-Eliashberg equations. Ref.~\cite{Wang:2016hir} showed that this ends up being equivalent to the RG approach under a certain local approximation. We will now briefly argue that this equivalence extends all the way to $d=2$.

Introducing a gap term $\t \Delta(\omega) \psi \psi+h.c.$ in the Hamiltonian, linearizing around $\t \Delta=0$ at the onset of the instability and assuming an even profile, the gap equation reads~\cite{Wang:2016hir}
\be
\t \Delta(\omega)= \frac{1}{2N}\int_{0}^\infty d\omega' u(\omega,\omega')\frac{\t \Delta(\omega')}{A(\omega')}\label{lineardeltaET}
\ee
where $A(\omega)= \omega + \Lambda^{1/3}\omega^{2/3}$ and the convolution kernel is 
\be
u(\omega,\omega')= \frac{2\Lambda^{1/3}}{3} \left(\frac{1}{|\omega-\omega'|^{1/3}}+\frac{1}{|\omega+\omega'|^{1/3}}\right) \,.
\ee
The above equation can be solved numerically, thus confirming the critical value \eqref{criticalN} for the existence of solutions. 

The integral equation can be transformed into a differential equation by means of a ``local approximation''
\be
 u(\omega, \omega') \approx \left\{ 
\begin{array}{ccc}
2u(\omega) & , & \omega'<\omega \\
2u(\omega') & , & \omega'>\omega
\end{array}
\right. \,,
\label{localapp}
\ee
whose validity we have checked by comparing with the full solution. Differentiating twice with respect to $\omega$ on both sides of the integral equation obtains
\be
\frac{d}{d\omega}\left(\frac{\t \Delta'(\omega)}{u'(\omega)}\right)=\frac{\t \Delta(\omega)}{N A(\omega)}\,.
\label{diffomega}
\ee

Physically, the local approximation means that non-local effects (in frequency space) are small. Then we can hope to match with the RG approach, which is intrinsically local. The explicit map is found to be
\be
\lambda(\omega) = f(\omega)\frac{\t \Delta(\omega)}{\t \Delta'(\omega)} \,\, , \,\, f(\omega)=\frac{8\pi}{3}\frac{u'(\omega)}{u(\omega)} \,.\label{SDEtoRG}
\ee
Rewriting \eqref{diffomega} as an equation for $\lambda'(\omega)$ and using that $\frac{d\log u(\omega)}{d\log\omega}=-1/3$,
 we end up with
\be
\frac{d\lambda(\omega)}{d\log\omega} = -\frac{8\pi}{9}+\frac{\lambda}{3}-\frac{\lambda^2}{4\pi N}\,,
\ee
which agrees with the RG beta function \eqref{BCSRG} evaluated at the critical point $\alpha=1/3$.\footnote{The function $f(\omega)$ is fixed by reprodicing the 1-loop renormalization term $\sim \lambda^2$.} Eq.~(\ref{SDEtoRG}) connects explicitly the formation of a gap (below which the gap function becomes stationary, $\t \Delta'(\omega)=0$), with the divergence $\lambda \to \infty$ of the BCS coupling.

%%%%%%%%%%%%%%%%%%%%%%%%%%%%
%%%%%%%%%%%%%%%%%%%%%%%%%%%%
%%%%%%%%%%%%%%%%%%%%%%%%%%%%
%%%%%%%%%%%%%%%%%%%%%%%%%%%%
\section{Discussion and future directions}\label{sec:concl}

In this work we have shown that soft bosonic modes, which are generic in non-Fermi liquid models in $d=2$ dimensions, lead to infrared divergences at finite temperature and to a breakdown of perturbation theory. We have argued that these divergences are resolved in terms of self-consistent boson and fermion self-energies. The $\phi^4$ interaction becomes dangerous irrelevant at finite temperature, and the self-consistent equation for the boson self-energy resums the corresponding bubble diagrams. This leads to a boson mass $m_b^2 \sim \lambda_\phi T$ that gaps the static mode; on-shell it is still irrelevant for the higher modes, which continue to have a $z_b=3$ scaling. We then used this as an input for the fermion sector, where we found a consistent solution to the IR problems by including all rainbow diagrams. This corrects the usual Eliashberg equation for the fermion self-energy, in agreement with~\cite{Wang:2017kab}. The main result is a thermal contribution $\Sigma_T \sim \sqrt{g^2 T}$ that dominates over a wide range of frequencies and temperatures. A similar scaling was recently found in~\cite{klein2020normal}. We argued that this modifies the standard picture of quantum phase transitions and the quantum to classical crossover in conceptually important ways, and we discussed some of the implications.

These thermal effects will have important consequences for the phenomenology of NFL models. To end we would like to discuss some of the future directions. 

First, it will be interesting to analyze how the thermal scale $\xi_T \sim (g^2T)^{-1/2}$ modifies thermodynamic and transport properties. Furthermore, while we have argued that superconductivity becomes irrelevant above $N_c=8$, we plan to present a detailed analysis at finite temperature in future work, finding the solutions to the corrected Eliashberg equations and revisiting the role of the first Matsubara frequencies~\cite{Wang:2017teb, 2019PhRvB..99n4512W, 2019arXiv191201797C}. Another direction recently explored in~\cite{klein2020normal, Xu:2020tvb} is the relevance of thermal effects to numerical quantum Monte Carlo results~\cite{Schattner2016, 2017PNAS..114.4905L, Berg2019, liu2019itinerant}, something that deserves further study both at finite but small temperature and including the effects of pairing interactions. Finally, the methods of this paper for a $z_b=3$ bosonic scaling could be extended to more general $z_b>1$. We hope to present this general analysis in future work.

\acknowledgements

We are grateful to A. Chubukov. A. Klein, S. Raghu and H. Wang for comments and discussions. We are supported by CNEA, Conicet (PIP grant 11220150100299), and UNCuyo.

\newpage

%%%%%%%%%%%%%%%%%%%%%%%%%
%%%%%%%%%%%%%%%%%%%%%%%%%
\appendix

\begin{widetext}

%%%%%%%%%%%%%%%%%%%%%%%%%%%%
%%%%%%%%%%%%%%%%%%%%%%%%%%%%
\section{One loop calculations}\label{app:calcs}

In this Appendix we calculate the one loop boson and fermion self-energies.

The 1-loop correction to Landau damping  is generated by bare one-loop fermion bubble diagrams.  We have
\bea
\Pi(\Omega_m, q)&=& \frac{g^2}{N} T \sum_n\,\int \frac{d^2 p}{(2\pi)^2}\,G(\omega_n, p) G(\omega_n+\Omega_m, p+q) \nonumber\\
&=& \frac{k_F g^2}{N}T\sum_n \int \frac{dp_\perp d\theta}{(2\pi)^2}\frac{1}{i\omega_n-v p_\perp}\frac{1}{i(\omega_n+\Omega_m)-v (p_\perp+q\cos\theta)}\\
&=& i \frac{k_F g^2}{v N}T\sum_n \int \frac{ d\theta}{2\pi}\frac{\Theta(\Omega_m+\omega_n)-\Theta(\omega_n)}{i\Omega_m-vq\cos\theta}\nonumber
\eea
The angular integral can be easily performed by going to the complex variable $z=e^{i\theta}$ and integrating on the unit circle, obtaining
\bea
\Pi(\Omega_m, q)&=& \frac{k_F g^2}{v N} T\sum_n \frac{\Theta(\Omega_m+\omega_n)-\Theta(\omega_n)}{\sqrt{(vq)^2+\Omega_m^2}}\nonumber\\
&=& \frac{k_F g^2}{v N} \frac{|\Omega_n|}{\sqrt{\Omega_n^2+(vq)^2}}\approx 
\frac{|\Omega_n|}{q}\,.
\eea
This is the result shown in equation (\ref{eq:PIFL}), that is independent of temperature.

The computation of quantum self-energy $\SigmaN(\omega)$ from expression \eqref{eq:SigmaNFL1} can be recast as
\bea
\SigmaN(\omega_n)&=&\Lambda^{1/3}(2\pi T)^{2/3}\frac13\sum_{m\neq n}\frac{{\rm sgn}(\omega_m)}{|m-n|^{1/3}}\\
&=&\Lambda^{1/3}(2\pi T)^{2/3}\frac{{\rm sgn}(\omega_n)}{3}\left(\sum_0^{\infty}\frac{1-\delta_{|n'|,m}}{|m-|n'||^{1/3}}-\sum_0^{\infty}\frac{1}{|m+|n'|+1|^{1/3}}\right)\,, \nonumber
\eea
with $|n'|=|n+1/2|-1/2$. % for $n\geq 0$ and $n'=n+1$ for $n<0$.
Using the definition of the generalized Riemann zeta function (Hurwitz zeta) $\zeta(s,a)=\sum_0^{\infty}(m+a)^{-s}$ ($\zeta(s)=\zeta(s,1)$), we have
\bea
\sum_{m =0}^\infty\frac{1-\delta_{|n'|,m}}{ |m - |n'||^{1/3}}&=& \sum_{m=0}^{|n'|-1} \frac{1}{|m-|n'||^{1/3}}+\zeta(1/3)\,,\nonumber \\
\sum_{m=0}^{|n'|-1} \frac{1}{|m-|n'||^{1/3}}&=& \zeta(1/3)-\zeta(1/3,|n'|+1)\,.\nonumber 
\eea
Putting all together, we finally obtain 
\be
\SigmaN(\omega_n) = (2\pi T)^{2/3}\Lambda^{1/3}{\rm sgn}(\omega_n) \frac{2}{3} \left( \zeta(1/3)-\zeta(1/3,|n+1/2|+1/2)\right)\,,
\ee
which is the result shown in equation (\ref{eq:Rzeta1}) in the main text.

%%%%%%%%%%%%%%%%%%%%%%%%%%%%%
%%%%%%%%%%%%%%%%%%%%%%%%%%%%%
\section{Self-consistent boson mas}\label{app:mb}

In Sec.~\ref{subsec:mb}, the self-consistent boson mass was obtained in the effective theory of the static $\Omega_n=0$ mode, valid up to a cutoff $q^3<2\pi M_D^2 T$. Here we rederive this result by working in the full theory that includes higher Matsubara modes.

We introduce a cutoff $|\Omega_m|<\Lambda=(2\pi T)\Lambda'$ and subsequently take $\Lambda\to\infty$. It is useful to introduce the variable $x=q^3(2\pi T M_D^2)^{-1}$, obtaining
\bea
m_b^2 &=& \frac{\lambda_\phi T}{2\pi}\sum_{m=-\Lambda'}^{\Lambda'} \int_0^\infty \frac{dq \, q}{q^2+m_b^2 +M_D^2\frac{|\Omega_m|}{q}}\nonumber \\
&=&\frac{\lambda_\phi T}{2\pi}\int_0^\infty \frac{dx}{x+a x^{1/3}} +\frac{\lambda_\phi T}{\pi}\sum_{m=1}^{\Lambda'}\int_0^\infty \frac{dx}{x+a x^{1/3}+m}\,.
\eea
Performing the sum in Matsubara frequencies and subsequently taking the divergent and finite parts as $\Lambda\to \infty$, obtains 
\bea
m_b^2\approx \frac{\lambda_\phi T}{2\pi}\int_0^\infty \frac{dx}{x+a x^{1/3}} +\frac{\lambda_\phi T}{\pi} \int_0^\infty  \left(\log\Lambda'-\psi(1+x+a x^{1/3})\right) + {\cal O}(\Lambda'^{-1})\,,
\eea
where $a= m_b^2 (2\pi TM_D^2)^{-2/3}$ and $\psi(x)$ denotes the digamma function. 

As the model is tunned to criticality, we need to subtract the $T=0$ contribution. As we will see, this is enough to cancel all divergences. The zero-temperature contribution can be suitably written as
\bea
m_0^2 &=& \frac{\lambda_\phi}{2\pi} \int_{-\Lambda}^{\Lambda}\frac{d\Omega}{2\pi} \int_0^\infty \frac{dq \, q}{q^2+M_D^2\frac{|\Omega_m|}{q}}\approx\frac{\lambda_\phi T}{\pi}\int_0^\infty \log\left(\frac{\Lambda'}{x}\right)\,.
\eea 
Subtracting the zero temperature contribution we can safely take $\Lambda'\to\infty$. Finally, in order to verify that the resulting expression is finite, we split the integration domain between $x<1$ and $x>1$. In the latter domain, we can neglect the $a x^{1/3}$ terms, thus having
\bea
m_b^2 -m_0^2 = &\,& \frac{\lambda_\phi T}{2\pi}\int_0^1 \frac{dx}{x+a x^{1/3}} 
+ \frac{\lambda_\phi T}{\pi} \int_0^1 dx \left(\log(x)-\psi(1+x+a x^{1/3})\right) \\
&+& \frac{\lambda_\phi T}{2\pi} \int_1^\infty dx\left(2\log(x)-2\psi(1+x)+\frac{1}{x}\right)\,. \nonumber
\eea
By the asymptotics of the digamma function $\psi(x)\approx \log(x) +(2x)^{-1}$, we note that the integral in the second line above is indeed finite and small. Moreover, the second term in the first line can be seen to be suppressed by powers of $a\sim M_D^{-4/3}$. Neglecting these terms we end up with 
\be
m_b^2 -m_0^2 \approx  \frac{\lambda_\phi T}{2\pi}\int_0^1 \frac{dx}{x+a x^{1/3}}=  \frac{\lambda_\phi T}{2\pi} \int_{q^3<2\pi T M_D^2} \frac{dq \, q}{q^2+m_b^2}\,,
\ee
which reproduces the result (\ref{eq:mb-self}).

%%%%%%%%%%%%%%%%%%%%%%%%%%%%%
%%%%%%%%%%%%%%%%%%%%%%%%%%%%%
%%%%%%%%%%%%%%%%%%%%%%%%%%%%%
\section{Running BCS coupling}\label{runningBCS}

In this appendix we derive the solutions to \eqref{BCSRG}, evaluated on the critical point $\alpha=1/3$, and discuss the BKT behavior. 
As we are interested in physics near the critical value of $N=N_c=8$, let's replace $N=8+\delta$ and consider both signs for $\delta$.  For a putative value $\lambda_0$ (which we assume small) of the BCS coupling at the UV cutoff $M_D$, the solution reads
\be
\lambda(\mu)= \frac{2\pi}{3}\left(8+\delta-\sqrt{\delta(8+\delta)}{\rm tanh}\left({\rm coth}^{-1}\left(\frac{2\pi\sqrt{\delta(8+\delta)}}{2\pi(8+\delta)-3\lambda_0}\right)+\frac16\sqrt{\frac{\delta}{8+\delta}}\log\frac{M_D}{\mu}\right)\right)
\ee  
For $\delta>0$, the above solution flows to a finite value, corresponding to the fixed point~\eqref{fixedBCS}.

For $\delta<0$, the solution satisfying the appropriate initial condition now can be written as 
\be
\lambda(\mu)= \frac{2\pi}{3}\left(8-|\delta| + \sqrt{|\delta|(8-|\delta|)}{\rm tan}\left(-{\rm cot}^{-1}\left(\frac{2\pi\sqrt{|\delta|(8-|\delta|)}}{2\pi(8-|\delta|)-3\lambda_0}\right)+\frac16\sqrt{\frac{|\delta|}{8-|\delta|}}\log\frac{M_D}{\mu}\right)\right)
\ee  
Since $\tan(x)$ diverges for $x=(n+1/2)\pi$, the effective coupling diverges at a finite value of the energy scale $\mu$. Near the critical value, {\it i.e.} for $|\delta|\to 0$, the largest scale at which this occurs is
\be
\frac{\mu}{M_D} \approx e^{-\frac{6\pi\sqrt{8-|\delta|}}{\sqrt{\delta}}}\,.
\ee  
This scale sets the magnitude of the physical order parameter at the onset of the phase transition, agrees with the correlation length (\ref{eq:xi}), and reveals the BKT scaling.

\end{widetext}

\bibliography{NFL}{}
\bibliographystyle{utphys}
\end{document}